\documentclass[12pt]{article} 

\usepackage{url,hyperref,lineno,microtype,subcaption}
\usepackage[onehalfspacing]{setspace}

\usepackage{amsmath,amsfonts,amssymb,graphicx,amsthm,geometry}
\usepackage{stackrel,stmaryrd,mathabx,etoolbox,framed,accents}
\usepackage[all]{xy}

\def\be{\begin{equation}}
\def\ee{\end{equation}}
\def\ba{\begin{eqnarray}}
\def\ea{\end{eqnarray}}

\makeatletter
\newsavebox{\@brx}
\newcommand{\llangle}[1][]{\savebox{\@brx}{\(\m@th{#1\langle}\)}%
  \mathopen{\copy\@brx\kern-0.5\wd\@brx\usebox{\@brx}}}
\newcommand{\rrangle}[1][]{\savebox{\@brx}{\(\m@th{#1\rangle}\)}%
  \mathclose{\copy\@brx\kern-0.5\wd\@brx\usebox{\@brx}}}
\makeatother

\newgeometry{vmargin={25mm}, hmargin={22mm,22mm}} 

\begin{document}

\setcounter{tocdepth}{1}

\title{Geometric tool kit for higher spin gravity (part I):\\ An introduction to the geometry of differential operators 
} 

\author{Xavier Bekaert}

\date{Institut Denis Poisson, Unit\'e Mixte de Recherche $7013$ du CNRS\\
Universit\'e de Tours, Universit\'e d'Orl\'eans\\
Parc de Grandmont, 37200 Tours, France\\
\vspace{2mm}
{\tt xavier.bekaert@lmpt.univ-tours.fr}
}

\maketitle

\vspace{5mm}

\begin{abstract}
These notes provide an introduction to the algebra and geometry of differential operators and jet bundles. Their point of view is guided by the leitmotiv that higher-spin gravity theories call for higher-order generalisations of Lie derivatives and diffeomorphisms. 
Nevertheless, the material covered here may be of general interest to anyone working on topics where geometrical (coordinate-free, global, generic) and mathematically rigorous definitions of differential operators are required.
\end{abstract}

\thispagestyle{empty}

\pagebreak

\setcounter{tocdepth}{3}

\pagestyle{empty}

\tableofcontents

\pagebreak

\pagestyle{plain}

\setcounter{page}{1}

\section{Introduction}

The recurrent theme behind the higher-spin extension of usual spacetime symmetries can be summarised as follows: \textit{Replace everywhere possible ``vector fields'' by ``differential operators''}. For instance, a by now standard recipe for constructing the higher-spin generalisation of a spacetime symmetry algebra (say of isometries or conformal transformations) is by allowing for higher-derivative differential operators as extra symmetry generators beyond vector fields. 
More precisely, higher-spin algebras of rigid symmetries are higher-order extensions of the isometry (or conformal) algebra obtained by taking its enveloping algebra (with respect to some irreducible representation). The gauging, \`a la Cartan, of such higher-spin algebras leads to higher-spin gravity theories.\footnote{Many pedagogical reviews on higher-spin gravity are available by now: advanced ones \cite{Reviews} as well as introductory ones \cite{Introductions}. Two books of conference proceedings also offer a panorama of this research area \cite{Proceedings}.} Furthermore, another standard recipe in higher-spin gravity is to pack the tower of metric-like fields of all ranks into a single generating function on the cotangent bundle, which can be interpreted as the symbol of a differential operator. Accordingly, higher-spin gravity calls for higher-order generalisations of Lie derivatives and diffeomorphisms.
This issue motivates this introduction to the mathematics (algebra and geometry) of differential operators.

Coordinate-free formulations of jet bundles\footnote{A classical textbook on the subject is \cite{Saunders}. Other textbooks covering the geometry of jet bundles are \cite{Olver,Krasilshchik,Sardanashvily1} while shorter introductions are \cite{Vinogradov}.} and differential operators\footnote{An inspiring source on the algebro-geometric approach to smooth manifolds, differential operators and jet bundles is the classical textbook \cite{Nestruev}. A concise pedagogical introduction is \cite{Sardanashvily2}.} are reviewed here in a self-contained way. The leitmotiv behind these notes is to view differential operators (respectively, jet fields) as natural higher-derivative generalisations of vector fields (respectively, differential one-forms). Although this material is standard, the way some notions are presented and some side observations might be original, to the best of the author's knowledge.

\subsection{Plan}

The plan of this paper is as follows:

Section \ref{motivation} initiates these notes with a general outlook on differential operators as natural generalisations of vector fields. Pragmatic readers may jump directly to Section \ref{0-orderstr} where the technical part is actually starting. 

The section \ref{0-orderstr} aims to immediately introduce the reader to the modern algebraic approach to geometry where one shifts the focus from the manifold itself to the algebra of functions on it. In particular, the one-to-one correspondence between the points of a manifold and the maximal ideals (or, dually, the multiplicative functionals) of the corresponding algebra of functions is reviewed. This abstract detour proves to be necessary in order to address the geometry of differential operators later on.

Similarly, the modern definition of vector fields on a manifold as derivations of the algebra of functions is adopted in Section \ref{tgtvects}. The one-to-one correspondence between diffeomorphisms and automorphisms of the algebra of functions is also discussed. The (co)tangent bundles are adressed along similar lines and the structure of symplectic (respectively, Schouten) algebra for the space of functions on the cotangent bundle (respectively, symbols of differential operators) is presented.

First-order differential operators and the algebraic structure they span are introduced in Section \ref{diffopss}. This section prepares the ground for the transition from first-order to higher-order differential structures in the next section. First-order (co)jets are also introduced.

Section \ref{co/jets} is the main goal of these notes. There, jets (respectively, cojets) are introduced as higher-order generalisations of cotangent (respectively, tangent) vectors. Various standard definitions of differential operators are reviewed such as the elegant algebraic definition due to Grothendieck. In order gain some geometrical intuition and stress the analogy with vector fields, a definition of differential operators as sections of cojet bundles is presented. Finally, the almost-commutative structure of the algebra of differential operators is emphasised.

\subsection{Teaser}

Wishfully thinking, higher-spin gravity might provide a fresh viewpoint on some modern geometrical notions perhaps less familiar to theoretical physicists, as well as a new playground for applications of these conceptual tools.  In order to proceed step by step, most of these notes will be devoted to an introduction for physicists to some modern (though, nowadays, textbook) viewpoints and developments in differential geometry, some of which could turn out to be useful in higher-spin gravity. 

More specifically, these lecture notes will be divided into four parts providing, respectively, introductions to the following subjects:
\begin{itemize}
	\item[$\bullet$] [Part I] the geometry of differential operators
	\item[$\bullet$] [Part II] the universal enveloping algebras of Lie algebroids
	\item[$\bullet$] [Part III] the general theory of connections on fibre bundles 
	\item[$\bullet$] [Part IV] higher-order frames and jet groups
\end{itemize}

\subsection{Disclaimer}

A first disclaimer is that these lecture notes are not intended for the pragmatists. They will be disappointed by the content since the material reviewed here does not necessarily provide tools for performing new calculations. In fact, a second disclaimer is that, despite its title, these lecture notes will not review nor address any practical problem in higher-spin gravity. The hope behind these lecture notes is rather that some better understanding of the geometry of higher-order differential operators and the corresponding potential generalisations of connections might contribute to foster some advances in higher-spin gravity. For instance, the subtleties encountered around the proper functional classes of infinitesimal vs finite automorphisms of the algebra of differential operators might have to do with difficulties faced by higher-spin interactions, such as their elusive (non-)locality properties. Also, the non-tensorial nature of cojets might be at the root of the problem of minimal coupling higher-spin gauge fields to gravitational backgrounds, in the metric-like formulation.

\pagebreak

\section{Enveloping geometry: Differential operators as higher vector fields}\label{motivation}

To start these notes, let us summarise its main emphasis and viewpoint in non-technical terms.

As is well-known, one of the leitmotives of modern theoretical physics is that fundamental interactions are governed by symmetry principles. Accordingly, one will try to motivate via symmetries the relevance of some higher-order structures on a manifold, such as the (co)jets that generalise the (co)tangent vectors and probe the differential structure beyond first-order. More specifically, higher-order differential operators will be introduced as natural generalisations (or better, as refinements, in the sense that they probe further the smooth structure of the manifold) of vector fields seen as infinitesimal symmetries of a smooth manifold.

Geometric symmetries, \textit{i.e.} finite or infinitesimal symmetries of a smooth manifold such as diffeomorphisms or vector fields, have an algebraic counterpart as automorphisms or, respectively, derivations of the algebra of functions on the (possibly fibred) manifold, \textit{i.e.} linear transformations which are compatible with the pointwise product structure of this algebra. 
Indeed, it is well known that vector fields can be thought of, \textit{geometrically}, as infinitesimal diffeomorphisms (\textit{i.e.} infinitesimal symmetries of the manifold) or, \textit{algebraically}, as derivations (\textit{i.e.} infinitesimal symmetries of the algebra of functions on the manifold).
More generally, the notion of infinitesimal symmetries of a fibred manifold is nowadays captured at the algebraic level by Lie-Rinehart algebras (\textit{e.g.} the algebra of first-order differential operators) and at the geometrical level by Lie algebroids, \textit{i.e.} vector bundles endowed with an anchor and a Lie bracket on their space of sections (a paradigmatic example being the tangent bundle whose sections are precisely the vector fields).

In field theory, a standard generalisation of geometrical symmetries are projective representations (the transformation of a function comes not only from a transformation of the coordinates but also with an extra multiplication by a factor) that correspond infinitesimally to first-order differential operators (sums of a vector field plus a function). The latter can be seen as endomorphisms (\textit{i.e.} linear transformations) compatible with the first-order structure of the space of functions. However, from this algebraic perspective there is no compelling reason to stop at first order if one considers smooth functions. In this sense, higher-order differential operators can be thought of as infinitesimal symmetries of the space of functions probing further the smooth structure of the manifold. 

There is actually a systematic procedure for performing this higher-order generalisation: passing to the universal enveloping algebra of the corresponding Lie-Rinehart algebra of infinitesimal symmetries.
For instance, the associative algebra of differential operators ($\Leftrightarrow$ infinitesimal higher-order symmetries) arises as the universal enveloping algebra of the Lie algebra of vector fields ($\Leftrightarrow$ infinitesimal diffeomorphisms). In more physical terms, the algebra of differential operators can be seen as a non-commutative algebra of functions on the quantum phase space, whose classical limit is a commutative algebra of functions on phase space. In more mathematical terms,  the associative algebra of differential operators is called ``almost-commutative'', in the sense that the product of two differential operators has order strictly higher than their commutator. In other words, differential operators commute modulo lower order terms. 

The various steps of increasing abstraction that one performs when passing from the manifold itself to its almost-commutative algebra of differential operators can be summarised as follows:

\begin{center}
Manifold $\stackrel{\text{duality}}{\Longleftrightarrow}$ Commutative algebra of functions 
\end{center}
\begin{center}
$\stackrel{\text{derivations}}{\Longleftrightarrow}$ Lie-Rinehart algebra of 1st-order differential operators
\end{center}
\begin{center}
$\stackrel{\text{enveloping}}{\Longleftrightarrow}$ Almost-commutative algebra of differential operators
\end{center}

\noindent The arrow ``duality'' illustrates the modern algebraic view on geometry where a manifold is equivalently described in terms of the commutative algebra of functions on this manifold. The arrow ``derivations'' stands for the addition of vector fields to the previous algebra of functions, thereby leading to the vector space of 1st-order differential operators endowed with a structure of Lie algebra via the commutator. The arrow ``enveloping'' corresponds to the step where one removes the bound on the number of derivatives, thereby obtaining the associative algebra of  differential operators.

But one can also proceed backward and reconstruct step by step the manifold itself from its almost-commutative algebra of differential operators by considering subalgebras as follows:

\begin{center}
Almost-commutative algebra of differential operators
\end{center}
\begin{center}
$\stackrel{\text{\small 1st-order sub.}}{\Longleftrightarrow}$ Lie-Rinehart algebra of 1st-order differential operators
\end{center}
\begin{center}
$\stackrel{\text{\small 0th-order sub.}}{\Longleftrightarrow}$ Commutative algebra of functions 
\end{center}
\begin{center}
$\stackrel{\text{\small max. ideals}}{\Longleftrightarrow}$ Manifold
\end{center}

However, there is a price to pay to interpret differential operators (with derivatives higher than one) as infinitesimal symmetries of the algebra of functions: one should relax the compatibility with the pointwise product, in so far as higher-order differential operators are \textit{not} derivations of the commutative algebra of functions. Nevertheless, in a sense they preserve its module structure. Still, one may  consider the commutative algebra of functions to be just the tip of the iceberg, a small subalgebra of the almost-commutative algebra of differential operators. While it is by now standard to admit that all the geometric information about a manifold is equivalently encoded into the algebraic properties of its commutative algebra of functions (on the manifold), one might go one step further and stress that it is also encoded into the almost-commutative algebra of differential operators on the manifold, \textit{i.e.} to adopt the following point of view: 

\ba
\text{Manifold}
&\Longleftrightarrow&\text{Commutative algebra of functions}\nonumber\\  
&\Longleftrightarrow&\text{Lie-Rinehart algebra of vector fields}\label{equivalence}\\
&\Longleftrightarrow&\text{Almost-commutative algebra of differential operators}\nonumber
\ea

For the sake of simplicity, rather than looking at finite automorphisms one may rather start by focusing on the infinitesimal automorphisms of the algebras, \textit{i.e.} their derivations.
On the one hand, all derivations of the commutative algebra of functions on a manifold are outer and provide an algebraic definition of vector fields on the manifold. On the other hand, all derivations of the Lie algebra of vector fields are inner. The Lie bracket of two vector fields identifies with their commutator and defines the Lie derivative of a vector field along another one.
Analogously, all derivations of the almost-commutative algebra of differential operators are inner, at least locally. 
In this sense, the adjoint action of any differential operator on the associative algebra of differential operators provides a natural higher-order generalisation of the Lie derivative along a vector field.

However, as can be expected from the equivalence between all the above notions (in a sense made mathematically precise via the generalised Milnor exercises reviewed in the following sections), the almost-commutative algebra of differential operators must have (essentially) the same collection of symmetries as the commutative algebra of functions. Locally, the automorphisms of the almost-commutative algebra of differential operators are indeed in one-to-one correspendence with the automorphisms of the commutative algebra of functions. Moreover, diffeomorphisms are (exactly) in one-one-correspondence with automorphisms of the commutative algebra of functions. This means that the automorphisms of the almost-commutative algebra of differential operators remain too narrow to look for a higher-order generalisation of diffeomorphisms.

Pursuing the above analogy and motivated by higher-spin gravity, one may consider somewhat larger classes of automorphisms that are called higher-spin diffeomorphisms. In more physical terms, higher-spin diffeomeorphisms can be thought of as the quantum version of canonical transformations.\footnote{Let us stress that the word ``quantum'' should be taken in a mathematical technical sense, not in a physical literal sense. As in deformation quantisation, ``quantum'' must be understood here as synonymous of ``associative'' while ``classical'' is synonymous of ``Poisson''.} Indeed, in a sense higher-spin diffeomeorphisms generalise the usual diffeomorphisms of a manifold in much the same way that canonical transformations generalise the diffeomorphisms of the configuration space in classical mechanics.
Turning back to differential operators, let us repeat that they can be seen as derivations of the almost-commutative algebra of functions on the quantum phase space. Since the derivations of commutative algebras of functions can be interpreted as vector fields on the corresponding manifold, one may in this sense assert that differential operators are nothing but some specific class of Hamiltonian vector fields on the quantum phase space. Moreover, differential operators are \textit{infinitesimal} higher-spin diffeomorphisms, so with this terminology in mind one might also view them as higher-spin vector fields. These side comments may be taken as possible justifications for the provocative subtitle of this section (``Differential operators as higher vector fields'') since there are several ways in which differential operators can be seen as generalisations of vector fields. 

This whole philosophical discussion can be summarised in the table \ref{analogysymmetries}.

\begin{table}
\begin{center}
\scriptsize
\begin{tabular}{
|c|c|c|c|c|c|}
 \hline
Geometry  & Manifold & Zeroth-order & First-order & Higher-order & Infinite-order\\ 
  & (smooth) & structure & structure & structure & structure \\ 
\hline
Algebra  & Maximal & Commutative & Lie-Rinehart & Almost-comm. & Associative \\ 
  & spectrum & algebra & algebra & algebra & algebra \\ 
\hline
Elements & Points & Functions & Vector fields & Differential & Beyond diff.\\ 
&&&& operators& operators\\ 
\hline
Infinitesimal & Infinitesimal & Vector & 1st-order & Differential & Beyond diff.\\
automorphisms & diffeos & fields & diff. ops & operators & operators \\ 
\hline
Automorphisms & Diffeos & Diffeos & Diffeos & Diffeos & Higher-spin\\ 
&&&&& diffeos \\\hline
\end{tabular}
\end{center}
\caption{Algebraic/geometrical structures and their symmetries}
\label{analogysymmetries}
\end{table}

\pagebreak 

\section{Differential structures of order zero: points and zeroth-order (co)jets}\label{0-orderstr}

To any manifold is associated the algebra of functions on it. But the converse is also true, in the sense that the manifold can sometimes be defined purely in terms of its algebra of functions, as the infinite family of maximal ideals of this commutative algebra.

Indeed, a possible definition of a point is as a maximal ideal of the algebra of functions on the manifold.
This definition is, unfortunately, very abstract. But it has the virtues of being at the same time purely algebraic and coordinate-free. This point of view proves to be fruitful to address the geometry of differential operators who lack some intuitive support (contrarily to vector fields, which one can always think as infinitesimal flows or as vectors tangent to a congruence of parametrised curves).

\subsection{Maximal ideal: What's the point?}

In order to get a flavor of how this abstract definition of a point is equivalent to more intuitive ones, 
consider a smooth manifold $M$ of finite dimension $n$. 
Associated with this manifold $M$ comes the commutative algebra $C^\infty(M)$ of smooth functions on $M$ taking values in a field $\mathbb K$ (in practice $\mathbb R$ here, but one may also replace it by $\mathbb C$). The pointwise product will be denoted by $\cdot$ and is defined by the equality
\be\label{pointwiseproduct}
(f\cdot g)|_m=f|_m\,g|_m\,,\qquad \forall m\in M\,,
\ee
where 
$f,g:M\to\mathbb K$ denote two functions on $M$ and the symbol $|_m$ stands for the evaluation at the point $m$.
When considered as a commutative algebra, the set $C^\infty(M)$ of smooth functions will often be called the \textbf{structure algebra}, because it encodes algebraically the geometric structure of the manifold.

Consider a given point $m\in M$. 
The commutative subalgebra 
\be
{\mathcal I}^0(m)\,:=\,\{\,f\in C^\infty(M)\,:\,f|_m=0\,\}
\ee
of functions $f$ vanishing at $m$ 
is an ideal of the structure algebra $C^\infty(M)$ that will be called the \textbf{contact ideal of order zero} at $m$. Indeed, the pointwise product $f\cdot g$ of a function $f\in {\mathcal I}^0(m)$ vanishing at $m$ with any function $g\in C^\infty(M)$ 
is an element $f\cdot g\in{\mathcal I}^0(m)$, since $(f\cdot g)|_m\,=\,f|_m\, g|_m\,=\,0$\,. 

The quotient 
\be\label{J0mM}
J^0_mM\,:=\,C^\infty(M)\,/\,{\mathcal I}^0(m)
\ee
is called the \textbf{zeroth-order jet space} at the point $m$.
By definition, it is the algebra of functions on $M$ quotiented by the equivalence relation: $f\sim g \Leftrightarrow f-g\in {\mathcal I}^0(m)\,$. 
In other words, two functions $f$ and $g$ are equivalent if and only if they take the same value at $m$: 
$f\sim g \Leftrightarrow f|_m=g|_m \,$. 
Such an equivalence class $[f]\in J^0_mM$, 
is called a \textbf{jet of order zero} or \textbf{zeroth-order jet} (or simply $0$-jet) at $m$.
The equivalence class $[f]$ is completely determined by the value of $f$ at $m$, so it is simply determined
by a number $f|_m\in\mathbb K$. Therefore,
the $0$-jet space \eqref{J0mM} at $m$ is one-dimensional and isomorphic to $\mathbb K$.
The field $\mathbb K$ is a vector space of dimension one over itself. Therefore the contact ideal ${\mathcal I}^0(m)$ is of codimension one inside ${\mathcal C}^\infty(M)$ since the quotient \eqref{J0mM} is of dimension one.
Moreover, a standard theorem of abstract algebra \cite[Chap.1]{Atiyah} states that a proper ideal of a commutative algebra is maximal iff the quotient of the commutative algebra by this ideal is isomorphic to a field.
Therefore, the isomorphism $J^0_mM\cong\mathbb K$, between the quotient of the commutative algebra $C^\infty(M)$ by the contact ideal ${\mathcal I}^0(m)$ and the field $\mathbb K$, shows that the contact ideal ${\mathcal I}^0(m)$ is of codimension one and maximal in $C^\infty(M)$.

Retrospectively, the \textbf{point} $m\in M$ can be identified with the maximal (or, equivalentely, codimension-one) ideal ${\mathcal I}^0(m)$ of the commutative algebra $C^\infty(M)$ of functions on the manifold. The collection of all maximal ideals of a commutative algebra $\mathcal A$ is called the \textbf{spectrum of maximal ideals} of this commutative algebra and will be denoted $\mathfrak{m}({\mathcal A})$.
For a compact manifold, the spectrum $\mathfrak{m}\big(\, C^\infty(M)\,\big)\cong M$ of maximal ideals of the commutative algebra of smooth functions on $M$ provides an algebraic reconstruction of the manifold $M$.\footnote{For non-compact manifolds, extra ``points at infinity'' actually arise as maximal ideals, \textit{c.f.} the ``ghosts'' discussed in \cite[Chap.8]{Nestruev}.}

\vspace{3mm}
\noindent{\small\textbf{Example 1 (Polynomials)\,:} Consider the vector space ${\mathbb K}^n$ with Cartesian coordinates $(y^1,\cdots,y^n)$ and the commutative subalgebra ${\mathbb K}[y^1,\cdots,y^n]\subset C^\infty({\mathbb K}^n)$ of \textit{polynomial} functions on ${\mathbb K}^n$ where $\mathbb K$ is an algebraically closed field. By a theorem of Hilbert (often called the ``weak form of Nullstellensatz'', see \textit{e.g.} \cite[Ex.17,Chap.5]{Atiyah}), the maximal ideals ${\mathcal I}^0(x^1,\cdots,x^n)\subset{\mathbb K}[y^1,\cdots,y^n]$ are the zeroth-order contact ideals spanned by polynomials vanishing at $(y^1,\cdots,y^n)=(x^1,\cdots,x^n)$, \textit{i.e.}
\be
{\mathcal I}^0(x^1,\cdots,x^n)\,:=\,\{\,P(y^1,\cdots,y^n)=\sum\limits_{a=1}^n(y^a-x^a)\,P_a(y^1,\cdots,y^n)\,\}\,,
\ee
where $P_a$ ($a=1,\cdots, n$) are polynomials.
The spectrum $\mathfrak{m}(\,{\mathbb K}[y^1,\cdots,y^n]\,)$ of maximal ideals is thus isomorphic to ${\mathbb K}^n$, which should be seen as an affine space (indeed the origin is not singled out any more in this construction). In a coordinate-free and basis-independent way, one would consider a vector space $V$ of finite dimension $n\in \mathbb N$ and the symmetric algebra $\odot(V^*)$ over its dual (\textit{i.e.} the commutative algebra of totally symmetric multinear forms on $V$). The spectrum of its maximal ideals can be taken as a (rather abstract) definition of the \textit{affine} space $A$ modeled on the \textit{vector} space $V$, as would be done in algebraic geometry. The symmetric algebra $\odot(V^*)$ is then interpreted as the commutative algebra of polynomial functions on the affine space $A$. One recovers the previous construction by choosing a basis $\{\texttt{e}_a\}$ of $V$ and denoting the components of vectors in this basis as the Cartesian coordinates $y^a$ (\textit{i.e.} $y=y^a\texttt{e}_a\in V$). Then $V$ identifies with ${\mathbb K}^n$ and the symmetric algebra $\odot(V^*)$ of symmetric multilinear forms $\alpha$ over $V^*$,
\be\label{alphay}
\alpha(y)=\sum\limits_{r\geqslant 0} T_{a_1\cdots a_r}\texttt{e}^{*a_1}\odot\cdots\odot\texttt{e}^{*a_r}\,,
\ee
identifies with the commutative algebra ${\mathbb K}[y^1,\cdots,y^n]$ of polynomials
\be\label{Py}
P(y)=\sum\limits_{r\geqslant 0}T_{a_1\cdots a_r}\,y^{a_1}\cdots y^{a_r}\,.
\ee
}

\vspace{3mm}
From now on, one will treat explicitly the case ${\mathbb K}={\mathbb R}$ (although many statements hold for ${\mathbb K}={\mathbb C}$ as well). A commutative algebra over $\mathbb R$ with a unique maximal ideal will be called a \textbf{local algebra}. According to the algebraic geometry viewpoint, it corresponds to an algebra of real functions on a space made of a single point. Although this case may look somewhat exotic, later on it will appear repeatedly as fibres in jet bundles. A paradigmatic example of local algebra is the commutative algebra of formal power series. 

\vspace{3mm}
\noindent{\small\textbf{Example 2 (Formal power series)\,:} Consider the commutative algebra ${\mathbb R}\llbracket \varepsilon^1,\cdots,\varepsilon^n\rrbracket $ of formal power series in $n$ variables with real coefficients. In contrast with the algebra ${\mathbb R}[\varepsilon^1,\cdots,\varepsilon^n]$ of polynomials, the algebra ${\mathbb R}\llbracket \varepsilon^1,\cdots,\varepsilon^n\rrbracket $ of formal power series is a local algebra: its unique maximal ideal is the subalgebra $\mathfrak{m}(\,{\mathbb R}\llbracket \varepsilon^1,\cdots,\varepsilon^n\rrbracket \,)$ of formal power series with vanishing constant term (\textit{i.e.} with terms which are at least linear). This fact may look surprising at first sight, but the analogues of the zeroth-order contact ideals ${\mathcal I}^0(x^1,\cdots,x^n)$ with $(x^1,\cdots,x^n)\neq(0,\cdots,0)$
are \textit{not}\footnote{For instance, consider the case $n=1$ of formal power series in the single variable $\varepsilon$. The space ${\mathcal I}^0(1)$ of formal power series of the form $f(\varepsilon)=(\varepsilon-1)g(\varepsilon)$ with $g(\varepsilon)\in{\mathbb R}\llbracket \varepsilon\rrbracket $ (\textit{i.e.} $g(\varepsilon)=\sum_{r=0}^\infty g_r\,\varepsilon^r$)  is \text{not} a proper ideal in ${\mathbb R}\llbracket \varepsilon\rrbracket $. In fact, the formal power series $g(\varepsilon)=(1-\varepsilon)^{-1}=\sum_{r=0}^\infty \varepsilon^r\in {\mathbb R}\llbracket \varepsilon\rrbracket $ is such that $(\varepsilon-1)g(\varepsilon)=-1$, hence ${\mathcal I}^0(1)={\mathbb R}\llbracket \varepsilon\rrbracket $.} proper ideals of ${\mathbb R}\llbracket \varepsilon^1,\cdots,\varepsilon^n\rrbracket $. This is related to the fact all formal power series with non-vanishing constant term are invertible.
The algebra of formal power series can be defined in a coordinate-free way by considering a vector space $V$ of finite dimension $n\in \mathbb N$ and the commutative algebra of Taylor series at the origin of smooth functions on $V$. This commutative algebra will be denoted either $\overline{\odot}(V^*)$ (to stress that it is a completion of the symmetric algebra $\odot(V^*)$\,) or $J^\infty_0V$ (for consistency with notations to be introduced later). 
Two remarks are in order. The linear dual of the vector space $\odot(V)$ is isomorphic to the vector space mentioned above
\be
\overline{\odot}(V^*)\,\cong\, \big(\odot (V)\,\big)^*\,.
\ee
The commutative algebra $\overline{\odot}(V^*)=J^\infty_0V$ is local, hence it must be interpreted as an algebra of functions over a single point: the origin of $V$.
Again, one recovers the previous construction by choosing a basis $\{\texttt{e}_a\}$ of $V$ and denoting the components of vectors in this basis as Cartesian coordinates $y^a$. Then the commutative algebra $\overline{\odot}(V^*)$ is spanned by linear forms over $\odot (V)$ taking the form \eqref{alphay} where infinite sums are allowed and identifies with the commutative algebra ${\mathbb K}\llbracket y^1,\cdots,y^n\rrbracket$ of formal power series taking the form \eqref{Py} where infinite sums are allowed.
}

\subsection{Zeroth-order jet bundle}

The \textbf{zeroth-order jet bundle} $J^0M=\bigcup_m J^0_m M $ is the reunion of all $0$-jet spaces. Let $x^\mu$ be local coordinates on 
$M$. Then local coordinates on 
$J^0M$ are $(x^\mu,\phi)$. Retrospectively, one may define the algebra of smooth functions on $M$ as
the space $\Gamma(J^0M)\cong  C^\infty(M)$
of global sections of the $0$-jet bundle $J^0M$.
Indeed a function $\phi$ can be identified with a global section of the bundle $J^0M$ since such a section is precisely specified by the values $\phi(x)$ at each point. 

The zeroth-order jet bundle $J^0M$ defined above is actually a trivial line bundle over $M$: 
\be J^0M\,\cong\,M \times {\mathbb R}\,.
\ee
Geometrically, the image $\sigma(M)\subset J^0M$ of a global section $\sigma\in \Gamma(J^0M)$ of the zeroth-order jet bundle is nothing but the graph of the corresponding function $\phi\in C^\infty(M)$. 

\subsection{Multiplicative functionals: Again, what's the point?}

In some sense, the dual notion of a maximal ideal ${\mathcal I}\in\mathfrak{m}({\mathcal A})$ of a commutative algebra $\mathcal A$ with unit\footnote{All associative algebras will be assumed to possess a unit element in this paper.} over a field $\mathbb K$ is a \textbf{multiplicative linear form}, \textit{i.e.} a morphism $\alpha:{\mathcal A}\to\mathbb K$ of commutative algebras, that is to say a form on $\mathcal A$ which (i) is linear and (ii) preserves the product,
\be\label{morph0cojet}
\alpha(\lambda_1\phi_1+\lambda_2\phi_2)=\lambda_1\alpha(\phi_1)+\lambda_2\alpha(\phi_2)\,,\qquad\alpha(\phi_1\cdot\phi_2)=\alpha(\phi_1)\alpha(\phi_2)
\ee
for any two elements $\phi_1$ and $\phi_2$ of $\mathcal A$ and any two scalars $\lambda_1$ and $\lambda_2$ of $\mathbb K$.  To be more precise, the kernel of such a multiplicative linear form is a maximal ideal of codimension-one. Moreover, the kernel determines uniquely the multiplicative linear form. These facts can be seen as follows.

\vspace{3mm}
\noindent{\small\textbf{Proof:} Firstly, the kernel of an algebra morphism is an ideal. In particular, the kernel of a multiplicative linear form is indeed an ideal.
Secondly, consider a nontrivial multiplicative linear form $\alpha\in\text{Hom}(\mathcal{A},\mathbb{K})$, in other words a non-zero element $\alpha\in \mathcal{A}^*$ of the linear dual of the algebra ${\mathcal A}$ which is also a morphism of commutative algebras, \textit{i.e.} it satisfies \eqref{morph0cojet}. Since $\alpha$ is nontrivial by assumption, it is clearly surjective, $\text{Im}\,\alpha=\mathbb K$. Therefore, its kernel $\text{Ker}\,\alpha\subset\mathcal A$ is an ideal of codimension-one (and hence maximal) since ${\mathcal A}\,/\,\text{Ker}\,\alpha\cong \text{Im}\,\alpha=\mathbb K$. Any codimension-one ideal is necessarily a maximal proper ideal.
Thirdly, let $1_{\mathcal A}\in\mathcal A$ denote the unit element. Since $\mathbb K$ is a field, the algebra morphism property of $\alpha$ implies that $\alpha(1_{\mathcal A})=1$ (or $\alpha=0$). Any element $\phi\in\mathcal A$ decomposes as a sum $\phi=\lambda\,1_{\mathcal A}\,+\,v$ where $\lambda\in\mathbb K$ and $v\in\text{Ker}\,\alpha$. Finally, one has $\alpha(\phi)=\lambda$. \qed}

\vspace{3mm}
When $\mathcal A$ can be interpreted ``geometrically'' as an algebra of $\mathbb K$-valued functions on a ``manifold'' whose ``points'' are the multiplicative linear forms on $\mathcal A$, then one can interpret the latter as the functional ``evaluation at the corresponding point''.\footnote{See \cite[Chap.3]{Nestruev} for a criterion on commutative algebras over $\mathbb{K}=\mathbb{R}$ to be geometric.} 

\subsection{Zeroth-order cojet bundle}

A linear form $\alpha: C^\infty(M)\to\mathbb K$ on the structure algebra $ C^\infty(M)$ whose kernel is the contact ideal ${\mathcal I}^0(m)$ at the point $m\in M$ will be called a \textbf{zeroth-order cojet} (or $0$-\textbf{cojet} for short) at $m\in M$. The one-dimensional space of zeroth-order cojets at $m$ will be denoted $D^0_mM$. The \textbf{zeroth-order cojet bundle} $D^0M=\bigcup_m D^0_m M $ is the reunion of all $0$-cojet spaces.

By definition, the zeroth-order cojets at $m$ are linear functionals (on the space of smooth functions) that vanish on the contact ideal ${\mathcal I}^0(m)$, so one can see them as generalised functions (aka distributions) on $M$ with support at $m$.
More explicitly, let 
\be
\delta_m\,:\, C^\infty(M)\to\mathbb R\,:\,f\mapsto f|_m
\ee
denote the functional ``\textbf{evaluation at the point}'' $m$, \textit{i.e.} the Dirac distribution at $m$  defined by $\langle\,\delta_m\,,\,f\,\rangle\,=\,f|_m$ for any $f\in C^\infty(M)$. This functional is multiplicative, \textit{i.e.} it is a morphism of commutative algebras. Indeed, the definition \eqref{pointwiseproduct} of the pointwise product of two functions $f$ and $g$ is precisely such that the evaluation at each point $m$ is an algebra morphism sending the product of functions to the product of real numbers.
Moreover, the kernel of $\delta_m$ is obviously the maximal ideal ${\mathcal I}^0(m)$. 
The space $D^0_mM$ of $0$-cojets at $m$ is isomorphic to the dual of the space 
\be
J^0_mM= C^\infty(M)/{\mathcal I}^0(m)\cong \mathbb R
\ee
of $0$-jets at $m$, that is to say 
\be
D^0_mM\cong (J^0_mM)^*\cong \mathbb{R}\,.
\ee
In fact, the space $D^0_mM$  can be thought of as the one-dimensional space spanned by the evaluation functional $\delta_m$. In this vector space, the Dirac distribution $\delta_m$ is singled out as the only multiplicative linear form among them (because there is a unique linear form $\alpha:J^0_mM\to\mathbb R$ such that $\alpha(1)=1$).
The set $\text{Hom}\big(\, C^\infty(M)\,,\,\mathbb{R}\,\big)$ of multiplicative linear forms on the commutative algebra of smooth functions on $M$ provides an algebraic reconstruction of the manifold $M$. In fact, there is a bijection
\be
\delta_\bullet\,:\,M\stackrel{\sim}{\to}\text{Hom}\big(\, C^\infty(M)\,,\,\mathbb{R}\,\big)\,:\,m\mapsto\delta_m
\ee
between the set of multiplicative linear forms on the structure algebra $C^\infty(M)$
and the set of points of a smooth manifold $M$. This fact is often called the ``Milnor exercise'' \cite[Problem 1-C]{Milnor}
(for a short proof, see \textit{e.g.} \cite[Corollary 35.9]{Kolar}).

\vspace{5mm}\begin{figure}[h!]
\begin{framed}
\begin{center}
\textbf{Algebraic vs geometric formulation of points}
\end{center}

\noindent
For compact manifolds, the following notions are equivalent:

\noindent$\bullet$ a point $m$ of a smooth manifold $M$,

\noindent$\bullet$ a multiplicative functional on the commutative algebra $ C^\infty(M)$ whose kernel is the contact ideal ${\mathcal I}^0(m)$ of order zero at $m$: the evaluation functional $\delta_m\in D^0_mM$

\noindent$\bullet$ a maximal ideal of the commutative algebra $ C^\infty(M)$: the contact ideal ${\mathcal I}^0(m)$ of order zero at $m$.

\noindent For smooth non-compact manifolds, only the first two notions are equivalent.
\vspace{3mm}
\end{framed}
\end{figure}

\subsection{Pullback of functions}\label{pullbackfcts}

Let $F:M\to N$ be a map from the manifold $M$ (source) to the manifold $N$ (target).

The \textbf{pullback of a function} $g\in  C^\infty(N)$ on the target by the smooth map $F:M\to N$ is the function on the source defined as the precomposition of $g$ by $F$
\be
F^*g:=g\circ F\in  C^\infty(M)\,.
\ee 
The \textbf{pullback by} $F$ is the following map between the corresponding structure algebras 
\be
F^*\,:\, C^\infty(N)\to  C^\infty(M)\,:\,g\mapsto g\circ F\,.
\ee
It is the dual of the map $F:M\to N$ in
the sense that it is an algebra homomorphism between the structure algebras of the corresponding manifolds with the direction of the arrow in $F^*$ reversed with respect to the arrow in $F$:
it is a map from the structure algebra $ C^\infty(N)$ of the target manifold to the structure algebra $C^\infty(M)$ of the source manifold. 

The pullback operation itself, 
\be
{}^*\,:\,\text{Hom}_{\text{Smooth Man.}}(M,N)\to \text{Hom}_{\text{Comm.Alg.}}\big(\,{\mathcal C}^\infty(N)\,,\,{\mathcal C}^\infty(M)\,\big) \,:\,F\mapsto F^*\,,
\ee
is an antihomomorphism (\textit{i.e.} it reverses the order of multiplication) from the associative algebra of maps between smooth manifolds to the associative algebra of morphisms between their structure algebras, $(F\circ\, G)^*=G^*\circ F^*$.
If $F$ is bijective (\textit{e.g.} a diffeomorphism) then $F^*$ also is. More precisely, $(F^{-1})^*=(F^*)^{-1}$ as can be checked from the previous property. More generally, if $F$ is surjective (\textit{e.g.} a fibration) then $F^*$ is injective.

Along this line, an algebraic definition of a diffeomorphism of a smooth manifold $M$ (or between two manifolds) is as an automorphism of its structure algebra $ C^\infty(M)$ (respectively, isomorphism) of their structure algebras.
This algebraic definition is justified by the following theorem, sometimes called  ``Milnor exercise'' with a slight abuse of terminology.\footnote{See \textit{e.g.} \cite{Grabowski:2003a} for a version of this theorem with extremely mild assumptions.}

\vspace{3mm}
\noindent{\textbf{Theorem (Grabowski)\,:} \textit{A map $\Phi: C^\infty(N)\stackrel{\sim}{\to} C^\infty(M)$ between two structure algebras is an isomorphism of commutative algebras iff it is the pullback of a diffeomorphism $F:M\stackrel{\sim}{\to} N$ between these two manifolds, \textit{i.e.} $\Phi=F^*$.
}}
\vspace{3mm}

Therefore, the group of geometric diffeomorphisms of the manifold $M$ and the group of automorphisms of its structure algebra $ C^\infty(M)$ are isomorphic, $Diff(M)\cong{Aut}(\, C^\infty(M)\,)$. The above theorem admits a generalisation where the algebra isomorphism (respectively, diffeomorphism) is replaced with an algebra homomorphism (respectively, smooth map), \textit{c.f.} \cite[Corollary 35.10]{Kolar}, for which the original Milnor exercise \cite[Problem 1-C]{Milnor} corresponds to the case when $M$ is a single point and ${\mathcal C}^\infty(M)=\mathbb R$.

\pagebreak

\section{Differential structures of order one: tangent (co)vectors}\label{tgtvects}

As we saw in the previous section, diffeomorphisms of a smooth manifold $M$ can be seen algebraically as automorphisms of the structure algebra $ C^\infty(M)$.
The infinitesimal automorphisms of a commutative algebra are its derivations. Vector fields are the infinitesimal generators of flows on a manifold, \textit{i.e.} one-parameter groups of diffeomorphisms.  Therefore, it is natural to consider the algebraic definition of vector fields on a manifold as derivations of the commutative algebra of functions.

\subsection{Tangent vector fields as derivations}

\subsubsection{Derivations}

Let $V$ be a vector space.
A linear map $D:V\to V$ is sometimes called an \textbf{endomorphism} of the vector space $V$. The space $\text{End}(V)$ of endomorphisms of $V$ is an associative algebra for the composition $\circ$\,.

Any associative algebra $\mathcal A$ with product $\star$ can be endowed with a Lie algebra structure
via the commutator $[\,\,\,\stackrel{\star}{,}\,\,\,]$ as Lie bracket, in which case it will be called the \textbf{commutator algebra} and denoted $\mathfrak{A}$ when it is necessary to distinguish it from its associative counterpart. 
In particular, the general linear algebra $\mathfrak{gl}(V):=\mathfrak{End}(V)$ is the space $\text{End}(V)$ of endomorphisms endowed with a structure of Lie algebra via the commutator $[\,\,\,\stackrel{\circ}{,}\,\,\,]$.

A \textbf{derivation} of an algebra $\mathcal A$ is a linear map $D:{\mathcal A}\to{\mathcal A}$ obeying to the Leibniz rule: 
\be
D(a\star b)\,=\,D(a)\star b\,+\,a\star D(b)\,,\qquad\forall a,b\in{\mathcal A}\,.
\ee
The subspace $\mathfrak{der}({\mathcal A})\subset\mathfrak{gl}({\mathcal A})$ of derivations is a Lie subalgebra of the general linear algebra of endomorphisms of the vector space $\mathcal A$.

All \textbf{representations of a Lie algebra} $\mathfrak{g}$ \textbf{on an algebra} $\mathcal A$ will be assumed to be morphisms
of Lie algebras from $\mathfrak{g}$ to $\mathfrak{der}({\mathcal A})$.
In particular, the \textbf{adjoint representation} of the commutator algebra of an associative algebra $\mathcal A$ is the morphism of Lie algebras 
\be\label{adjointrep}
ad\,:\,\mathfrak{A}\to\mathfrak{der}({\mathcal A})\,:\,a\mapsto ad_a
\ee
where
\be
ad_{a}(b)\,:=\,[\,a\,\stackrel{\star}{,}\,b\,]=a\star b-b\star a\,,
\ee 
is a morphism of Lie algebras from the commutator algebra $\mathfrak{A}$ to the Lie algebra $\mathfrak{der}({\mathcal A})$ of derivations, \textit{i.e.} 
\be
[\,ad_{a_1}\,\stackrel{\circ}{,}\,ad_{a_2}\,]=ad_{[\,a_1\,\stackrel{\star}{,}\,a_2\,]}\,.
\ee 
The image 
\be
\mathfrak{inn}({\mathcal A}):=ad_{\mathfrak A}\subset\mathfrak{der}({\mathcal A})
\ee of the adjoint representation \eqref{adjointrep} is called the \textbf{Lie subalgebra of inner derivations}. One has the chain of inclusions: 
\be
\mathfrak{inn}({\mathcal A})\subset\mathfrak{der}({\mathcal A})\subset\mathfrak{gl}({\mathcal A})\,.
\ee
To be more explicit, an inner derivation is a derivation $D\in \mathfrak{der}({\mathcal A})$ of the form $D=ad_a$ for some element $a\in\mathcal A$. All other derivations, \textit{i.e.} $D\in \mathfrak{der}({\mathcal A})$ but $D\notin ad_{\mathcal A}$, are called \textbf{outer derivations}. All nontrivial derivations of a commutative algebra $\mathcal A$ are outer. They can in fact be interpreted as vector fields on the spectrum $\mathfrak{m}({\mathcal A})$ of maximal ideals. 

\vspace{3mm}
\noindent{\small\textbf{Example (polynomial vs formal vector fields)\,:} Consider the affine (vs vector) space $A$ (respectively, $V$) isomorphic to  ${\mathbb R}^n$ with the Cartesian coordinates $(y^1,\cdots,y^n)$. The Lie algebra of derivations of the commutative algebra ${\mathcal A}={\mathbb R}[y^1,\cdots,y^n]$ (vs ${\mathcal A}={\mathbb R}\llbracket y^1,\cdots,y^n\rrbracket $\,) of polynomial (vs formal) functions on this affine space (respectively, at the origin of this vector space) is called the \textbf{Lie algebra of polynomial (vs formal) vector fields}. Its elements take the form $\hat{X}=X^a(y)\frac{\partial}{\partial y^a}$ with components $X^a(y)$ which are polynomial (vs formal power series), \textit{i.e.} $\mathfrak{der}({\mathcal A})\,\cong\,{\mathcal A}\otimes V$ as vector spaces, where $V$ is the vector space on which the affine space $A$ is modeled.}

\subsubsection{Vector fields on a manifold}

In purely algebraic and coordinate-free terms, a (tangent) \textbf{vector field} $\hat{X}$ on $M$ can be defined as a 
derivation of the commutative algebra $ C^\infty(M)$ of functions on $M$.\footnote{Vector fields are hatted in order to emphasise their algebraic definition as differential operators and anticipate their higher-order generalisation.} 
In more concrete terms, a vector field is a linear map 
\be
\hat{X}\,:\, C^\infty(M)\to  C^\infty(M)\,:\,f\mapsto \hat{X}[f]
\ee 
where $\hat{X}[f]$ denotes the action\footnote{This action will sometimes also be denoted by $\hat{X}f$ for short.} of the vector field $\hat{X}$ on the function $f$, which is such that
\be\label{vectfielder}
\hat{X}[f\cdot g]=\hat{X}[f]\cdot g+f\cdot\hat{X}[g]\,,\quad\forall f,g\in C^\infty(M)\,.
\ee
Alternatively, it will also be identified with the \textbf{Lie derivative of a function} $f$ \textbf{along the vector field} $\hat{X}$, in which case it is denoted 
\be
{\mathcal L}_{\hat{X}}f\,:=\,\hat{X}[f]\,.
\ee
In local coordinates, this reads:
\be\label{X[f]}
\big(\,\hat{X}[f]\,\big)(x)\,=\,X^\mu(x)\,\partial_\mu f(x)\,.
\ee
Accordingly, the local expression of vector fields is $\hat{X}\,=\,X^\mu(x)\,\partial_\mu$ where the collection of vector fields $e_\mu:=\partial_\mu$ will be called the \textbf{vector fields of a coordinate basis}.
The vector space spanned by vector fields is usually denoted $\mathfrak{X}(M)$, but it will often be denoted in this text by $\mathcal{T}(M)$ for consistency with other choices of notation. Nevertheless, when the space $\mathcal{T}(M)$ of vector fields is endowed with a structure of Lie algebra via the commutator bracket, it will be denoted $\mathfrak{X}(M)=\mathfrak{der}\big(\, C^\infty(M)\,\big)$ to emphasise its Lie algebra structure.
In fact, when vector fields are seen as derivations, the \textbf{Lie bracket between two vector fields} $\hat{X}$ and $\hat{Y}$ 
is simply a fancy name for their commutator $[\hat{X}\,\stackrel{\circ}{,}\,\hat{Y}]$ with respect to the composition product $\circ$. In this context, it will be denoted $[\hat{X}\,,\hat{Y}]$ for simplicity.

\subsubsection{Flow of a vector field}

A vector field $\hat{X}$ on the manifold $M$ is said complete if it generates an action of the additive group $\mathbb R$ on the structure algebra $ C^\infty(M)$, \textit{i.e.} a group morphism
\be
\exp(\bullet\,{\mathcal L}_{\hat{X}})\,:\,\mathbb{R}\to {Aut}(\, C^\infty(M)\,)\,:\,t\mapsto\exp(\,t\,{\mathcal L}_{\hat{X}}\,)\,,
\ee
where $\exp(\,\mathbb{R}\,{\mathcal L}_{\hat{X}})\subset{Aut}(\,{\mathcal C}^\infty(M)\,)$ denotes a one-parameter group  
of automorphisms $\exp(\,t\,{\mathcal L}_{\hat{X}})\in {Aut}(\,{\mathcal C}^\infty(M)\,)$ of the structure algebra ${\mathcal C}^\infty(M)$,
\be\label{autodiffeomorphisms}
\exp(\,t\,{\mathcal L}_{\hat{X}}\,)\,:\, C^\infty(M)\stackrel{\sim}{\to} C^\infty(M)\,:\,f\mapsto f_t=\exp(\,t\,{\mathcal L}_{\hat{X}}\,)\,[f]\,.
\ee 
Geometrically, these automorphisms \eqref{autodiffeomorphisms} of the structure algebra $ C^\infty(M)$ correspond to diffeomorphisms of the manifold $M$,
\be\label{diffeomeorphisms}
\exp(\,t\,\hat{X})\,:\,M\stackrel{\sim}{\to}M\,:\,x^\mu\mapsto x_t^{\prime\mu}(x)=\exp(\,t\,{\mathcal L}_{\hat{X}}\,)\,[x^\mu]\,\,,
\ee
where we identified a point with its coordinates in a patch in order to relate explicitly the two viewpoints \eqref{autodiffeomorphisms}-\eqref{diffeomeorphisms}.
Any complete vector field $\hat{X}$ defines an action of the additive group $\mathbb R$ on the manifold $M$, \textit{i.e.} a group morphism  
\be
\exp(\bullet\hat{X})\,:\,\mathbb{R}\to Diff(M)\,:\,t\mapsto\exp(\,t\hat{X})\,,
\ee
called the \textbf{(global) flow on the manifold} $M$ \textbf{generated by the (complete) vector field} $\hat{X}$.\footnote{For technical details and proofs related to geometric flows, see \textit{e.g.} \cite{Lee}.} Explicitly, the relation between these two equivalent views (algebraic vs geometric) of diffeomorphisms is the following relation between the change of functions versus the change of coordinates: $f_t(x^\mu)=f(x_t^\mu)$.

\vspace{3mm}
{\noindent\small\textbf{Remark 1:}
Note that, the Lie derivative has been removed in the notation \eqref{diffeomeorphisms} with respect to \eqref{autodiffeomorphisms}, in order to distinguish between these two (equivalent) formulations: the (algebraic) formulation as automorphisms $\exp(\,t\,{\mathcal L}_{\hat{X}})\in {Aut}(\, C^\infty(M)\,)$ of the structure algebra and the (geometric) formulation as diffeomorphisms $\exp(\,t\,\hat{X})\in Diff(M)$ of the manifold. However, one will often be sloppy in the sequel and use the same notation for these equivalent notions.}

\vspace{3mm}
{\noindent\small\textbf{Remark 2:} If the vector field $\hat{X}$ on $M$ is incomplete, then the one-parameter group of automorphisms that it generates is not defined for some values of the parameter $t\in\mathbb R$. In practice, this means that the corresponding trajectories go to infinity for some finite value of the parameter. 
Nevertheless, any vector field $\hat{X}\in\mathfrak{X}(M)$ is integrable to a local flow $\exp(\bullet\hat{X})\,:\,I\to Diff(N)$ defined for an open subset $I\subseteq\mathbb{R}$ (\textit{e.g.} an open interval $I=]a,b[$\,) and a submanifold $N\subseteq M$. 
In the sequel, the adjective ``complete'' will often be dropped since local considerations are implicitly understood.}

\subsubsection{Lie derivative of vector fields via infinitesimal flow and via adjoint action}

The algebraic point of view suggests a natural definition for the \textbf{Lie derivative of a vector field} $\hat{Y}$ 
\textbf{along another vector field} $\hat{X}$ (denoted by ${\mathcal L}_{\hat X} \hat{Y}$) through the Leibniz rule:
${\mathcal L}_{\hat X}( \hat{Y}f)= ({\mathcal L}_{\hat X} \hat{Y})f+\hat{Y}({\mathcal L}_{\hat X}f)$.
This leads to 
\be
({\mathcal L}_{\hat X} \hat{Y})f={\mathcal L}_{\hat X}( \hat{Y}f)-\hat{Y}({\mathcal L}_{\hat X}f)= {\hat X}(\hat{Y}f)-{\hat Y}(\hat{X}f)=({\hat X}\circ\hat{Y}-{\hat Y}\circ\hat{X})f\,
\ee 
so that the Lie derivative of a vector field $\hat{Y}$
along another vector field $\hat{X}$ is indeed equal to the Lie bracket:
${\mathcal L}_{\hat X} \hat{Y}=[\hat{X},\hat{Y}]$.

Another perspective on the Lie derivative of a vector field $\hat{Y}$ along a vector field $\hat{X}$ is that the flow generated by the vector field $\hat{X}$ induces a one-parameter group $Ad_{\exp t\hat{X}}\in {Aut}(\,{\mathfrak{X}}(M)\,)$ of automorphisms of the Lie algebra ${\mathfrak{X}}(M)$ of vector fields:
\be
Ad_{\exp t\hat{X}}\,:\,{\mathfrak{X}}(M)\stackrel{\sim}{\to}{\mathfrak{X}}(M)\,:\,\hat{Y}\mapsto \hat{Y}^{\hat{X}}_t=\exp(\,t\hat{X})\circ\hat{Y}\circ\exp(-t\hat{X})\,.
\ee
In fact, the infinitesimal transformation $\frac{d}{dt}(\hat{Y}^{\hat{X}}_t)|_{t=0}$
of a vector field $\hat{Y}$ induced by the flow generated by the vector field $\hat{X}$ is the geometric 
definition of the Lie derivative of $\hat{Y}$ along $\hat{X}$. It leads to ${\mathcal L}_{\hat X}\hat{Y}=\frac{d}{dt}(\hat{Y}_t)|_{t=0}=[\hat{X},\hat{Y}]$ as it should.
The expression for the components of the Lie derivative of the vector field $Y^\mu$ along the vector field $X^\nu$
reads
\be
 ({\mathcal L}_{\hat{X}}\hat{Y})^\mu=X^\nu\partial_\nu Y^\mu-Y^\nu\partial_\nu X^\mu\,.
\ee

The Lie derivative itself
\be
{\mathcal L}\,:\,{\mathfrak{X}}(M)\stackrel{\sim}{\to}\mathfrak{inn}\Big(\,{\mathfrak{X}}(M)\,\Big)\,:\,\hat{X}\mapsto {\mathcal L}_{\hat{X}}
\ee
is nothing but the adjoint representation of the Lie algebra ${\mathfrak{X}}(M)$ of vector fields on itself.
In particular, it is indeed a Lie algebra morphism in the sense that 
\be
 [{\mathcal L}_{\hat{X}}\,\stackrel{\circ}{,}\,{\mathcal L}_{\hat{Y}}]={\mathcal L}_{[\hat{X},\hat{Y}]}\,.
\ee
Although this algebraic definition of the Lie derivative ${\mathcal L}=ad$ on ${\mathfrak{X}}(M)$ may look somewhat tautological, let us stress that nevertheless the Lie derivatives ${\mathcal L}_{\hat{X}}\in\mathfrak{inn}\Big(\,{\mathfrak{X}}(M)\,\Big)$ along vector fields $\hat{X}\in\mathfrak{X}(M)$ are extremely important since they exhaust all infinitesimal symmetries of the Lie algebra of vector fields. More precisely, a classical theorem of Takens \cite{Takens} asserts that all derivation of the Lie algebra of vector fields are inner derivations: 
\be
\mathfrak{inn}\Big(\,{\mathfrak{X}}(M)\,\Big)=\mathfrak{der}\Big(\,{\mathfrak{X}}(M)\,\Big)\,.
\ee 

As one can see, when vector fields are seen as differential operators, 
the geometric notions of Lie bracket and Lie derivatives of vector fields are redundant with the commutator bracket and the adjoint representation.
However, one knows that Lie derivatives also have a geometrical interpretation as infinitesimal diffeomorphisms 
pulled back to the point of origin. The conceptual distinction between the Lie bracket and the commutator bracket may be pertinent in ordinary gravity theories such as general relativity due to its clear geometric roots. However the identification between Lie bracket and commutator bracket might suggest a generalisation of the infinitesimal symmetries of ordinary gravity to higher-spin gravity
where one should better leave some of our geometrical prejudices
(since it is not obvious what the analogue of the role played by diffeomorphisms in the general covariance principle should be in higher-spin gravity).

\subsection{Tangent vectors}

Abstractly, the \textbf{tangent space} $T_mM$ at a point $m$ of a manifold $M$ can be defined as the space of 
equivalence classes of vector fields, where two vector fields $\hat{X}$ and $\hat{Y}$ are equivalent if they produce the same result, at the point $m$, on any given function . An element of the tangent space $T_mM$ is called a tangent vector at $m$.
The equivalence class will be denoted $\hat{X}|_m$ and called the value of the vector field at the point $m$. One has
\be\label{equivreltgtvect}
\hat{X}|_m=\hat{Y}|_m\qquad\Longleftrightarrow\qquad \hat{X}[f]|_m=\hat{Y}[f]|_m\,,\quad\forall f\in C^\infty(M)\,.
\ee
Equivalently, a tangent vector $v$ at $m$ can be defined as a linear form $v:C^\infty(M)\to\mathbb R$ on the structure algebra $C^\infty(M)$ that obeys the Leibnitz rule
\be\label{tgtvectorder}
v[f\cdot g]\,=\,v[f]\, g|_m\,+\,f|_m\, v[g]\,,\quad\forall f,g\in C^\infty(M)\,.
\ee
The relation between the latter and the former definitions is that such a linear map takes the form $v=\hat{X}|_m=\delta_m\circ\hat{X}$ for some vector field $\hat{X}\in\mathfrak{der}\big(\, C^\infty(M)\,\big)=\mathfrak{X}(M)$, where $\delta_m:C^\infty(M)\to\mathbb R$ is the  evaluation functional at the point $m$. The equivalence relation in the first definition arises from the fact that the kernel of $\delta_m$ is the zeroth-order contact ideal ${\mathcal I}^0(m)$.

Let $e_\mu:=\partial_\mu$ be the coordinate basis vector fields in some local coordinates $x^\mu$.
Consider the $n$ linear forms $e_\mu|_m$ on $C^\infty(M)$
acting on any function $f$ as $e_\mu|_m f:=(\partial_\mu f)|_m$.
They can be interpreted as the \textbf{coordinate basis of the tangent space} $T_mM$ at $m$.
The point $m$ will sometimes be replaced with its coordinates $x$. 

As usual, one may thus define the \textbf{tangent bundle} $TM=\bigcup_m T_m M $ as the union of all tangent spaces.
Local coordinates of this bundle are $(x^\mu,X^\nu)$ where $X^\nu$ are the components of a tangent vector (in the coordinate basis located at the point of coordinates $x^\mu$). A change of coordinate $x^\mu\mapsto x^{\prime\mu}$ on the base $M$ induces the transformation law $X^\nu\mapsto X^{\prime\nu}=\frac{\partial x^{\prime\nu}}{\partial x^{\mu}}\,X^\mu$ in the fibre.
One can reinterpret the vector space $\mathfrak{X}(M)$ of vector fields as the vector space of global sections $\Gamma(TM)$ of the tangent bundle.

A classical theorem of Pursell and Shanks provides a Lie-algebraic analogue of the generalised Milnor exercise of Subsection \ref{pullbackfcts}, where the commutative algebra of functions is replaced with the Lie algebra of vector fields \cite{Pursell}.

\vspace{3mm}
\noindent{\textbf{Theorem (Pursell \& Shanks)\,:} \textit{A map $\Phi:{\mathfrak{X}}(M)\stackrel{\sim}{\to}{\mathfrak{X}}(N)$ between the Lie algebras of vector fields on two manifolds is an isomorphism of Lie algebras iff it is the pushforward of a diffeomorphism $F:M\stackrel{\sim}{\to}N$ between these two manifolds, \textit{i.e.} $\Phi=F_*$.}}
\vspace{3mm}

This means that smooth manifolds somehow realise the perfect dream of group theorists: they provide objects characterised uniquely by their symmetries. Indeed, two smooth manifolds are isomorphic iff they have the same algebra of infinitesimal symmetries (\textit{i.e.} vector fields).

\subsection{Cotangent vectors}

The dual space $T^*_mM$ of linear forms $\alpha$ on the tangent space $T_mM$ is the \textbf{cotangent space} at $m$. 
An element of the cotangent space $T^*_mM$ is called a \textbf{cotangent vector}, or \textbf{tangent covector}, or a \textbf{one-form}, at the point $m$. 
The \textbf{cotangent bundle} is the manifold $T^*M=\bigcup_m T^*_m M $ and it has local coordinates $(x^\mu,p_\nu)$ where a change of coordinate $x^\mu\mapsto x^{\prime\mu}$ on the base $M$ induces the transformation law $p_\mu\mapsto p'_\mu=\frac{\partial x^{\nu}}{\partial x^{\prime\mu}}\,p_\nu$ in the fibre.

The sections of the cotangent bundle are called  \textbf{differential one-forms} $\alpha(x)$ of $\Omega^1(M):=\Gamma(T^*M)$. Their  components take the form $\alpha_\mu(x)$ in
the \textbf{dual basis to the coordinate basis}, usually written $dx^\mu$, which is a basis of the cotangent space $T^*_mM$.
This standard notation is motivated by the definition of the \textbf{differential of functions} as
the linear map
\be\label{diffofafct}
d\,:\, C^\infty(M)\to\Omega^1(M)\,:\,f\mapsto df
\ee
where the \textbf{differential of the function} $f\in  C^\infty(M)$ is the differential one-form $df\in\Omega^1(M)$ defined via its action on any vector field $\hat{X}\in \mathfrak{X}(M)$, 
\be
	\langle\, df\,,\,\hat{X}\,\rangle := \hat{X}[f]\,.
	\label{dfX}
\ee
In local coordinates, the right-hand-side of \eqref{dfX} is given by \eqref{X[f]}. Therefore, $df=dx^\mu\partial_\mu f$ in the coordinate basis since the left-hand-side of \eqref{dfX} then reads
\be
\langle\, dx^\mu\,\partial_\mu f\,,\,X^\nu\, \partial_\nu\rangle\,=\,\partial_\mu f\, X^\nu\, \langle\, dx^\mu\,,\,\partial_\nu\rangle\, =\, X^\mu\,\partial_\mu f
\ee
by definition of the dual basis, $\langle\, dx^\mu\,,\,\partial_\nu\rangle=\delta^\mu_\nu$\,. 

\subsection{Tangent tensors}

We will focus on totally symmetric tensors for later purpose.

\subsubsection{Symmetric contravariant tensor fields as functions on the cotangent bundle}

Since the tangent space is a vector space one may also define in a natural way the bundle
of \textbf{symmetric contravariant tensors} $\odot\, TM=\bigcup_m\, \odot T_m M $ together with the space $\Gamma(\odot\, TM)$ of \textbf{symmetric contravariant tensor fields}. 
To indicate the rank, one adds an upper index, \textit{e.g.} the bundle of contravariant tensors of rank one is the bundle $\odot^1TM\cong TM$.
Local coordinates of the bundle $\odot^r TM$ are $(x^\mu,T^{\mu_1\ldots \mu_r})$ and the components of a symmetric tensor field
of rank $r$ are written as $T^{\mu_1\ldots \mu_r}(x)$ in the coordinate basis $\partial_{\mu_1}\odot \cdots\odot \partial_{\mu_r}$.
Therefore, local coordinates of the infinite-dimensional bundle $\odot TM$ are 
\be
(x^\mu,T,T^\mu,\ldots,T^{\mu_1\ldots \mu_r},\ldots\,)\,.
\ee

An $\mathbb N$-\textbf{graded} vector space $V$ is a collection $\{\,V_i\,\}_{i\in\mathbb N}$ of vector spaces such that $V_i\cap V_j=\{0\}$ for $i\neq j$ and $V=\oplus_{i\in\mathbb N} V_i$. An element of $V_i$ is called a \textbf{homogeneous element of grading} $i$.
An algebra $\mathcal A$ with product $\star$ is $\mathbb N$-graded if it is 
so as vector space and 
if its product $\star$ obeys ${\mathcal A}_i\star {\mathcal A}_j\subseteq {\mathcal A}_{i+j}$. Moreover, if the algebra possesses a unit element $1\in\mathcal A$, then one further requires that $1\in {\mathcal A}_0$.
The space $\Gamma(\odot TM)$ 
of symmetric contravariant tensor fields endowed with the symmetric product (inherited from the fibres) is a commutative
$\mathbb N$-graded algebra (with the rank of tensors as grading).

A function on the cotangent bundle which is a homogeneous polynomial of degree $r$ in the fibre will be called a \textbf{symbol of degree} $r$ \textbf{on the manifold} $M$. The commutative algebra of symbols will be denoted by ${\mathcal S}(M)\subset C^\infty(T^*M)$.
It is $\mathbb N$-graded by the polynomial degree in the Cartesian coordinates on the cotangent spaces, which will be called the \textbf{polynomial degree in the momenta}.
The symmetric contravariant tensor fields of rank $r$ can be identified with symbols of degree $r$. This is clear in coordinates via the identification of the coordinate basis $\partial_\mu$ with the fibre coordinates $p_\mu$ and of the symmetric product $\odot$ with the obvious product of the variables $p_\mu$, thus one has the isomorphism
$\Gamma(\odot TM)\cong{\mathcal S}(M)$ of $\mathbb N$-graded commutative algebras. Indeed, an element
$T=\sum\frac1{r!}\,T^{\mu_1\ldots \mu_r}(x)\,\partial_{\mu_1}\odot \cdots\odot \partial_{\mu_r}$
of $\odot\mathcal{T}(M)$ can be mapped to a function $T(x,p)=
\sum\frac1{r!}\,T^{\mu_1\ldots \mu_r}(x) \,p_{\mu_1} \cdots p_{\mu_r}$ on $T^*M$ which is a smooth function in $x$ and a
polynomial in $p$.
The symmetric product $\odot$ in $\Gamma(\odot TM)$ given by
\be
(T_1\odot T_2)^{\mu_1\ldots \mu_{r_1+r_2}}(x)\,=\,\frac{(r_1+r_2)!}{r_1!\,r_2!}\,T_1^{(\mu_1\ldots \mu_{r_1}}(x)\,T_2^{\mu_{r_1+1}\ldots \mu_{r_2})}(x)
\ee
is mapped in ${\mathcal S}(M)$ to the pointwise product of functions on the cotangent bundle. Throughout these notes, curved brackets over a set of indices denote complete symmetrisation over all this indices, with weight one, \textit{i.e.} $T^{(\mu_1\ldots \mu_r)}=T^{\mu_1\ldots \mu_r}$.

Via the Leibniz rule, one can deduce the Lie derivative of the symmetric contravariant tensor field $T=T^{\mu_1\ldots \mu_r} \partial_{\mu_1}\odot \cdots\odot \partial_{\mu_r}$ along the vector field $\hat{X}=X^\nu \partial_\nu$ from the expression of the Lie derivative of vector fields $\hat{Y}=Y^\nu \partial_\nu$. In components, it takes the form 
\begin{eqnarray}
 ({\mathcal L}_{\hat{X}}T)^{\mu_1\ldots \mu_r}&=&X^\nu\partial_\nu T^{\mu_1\ldots \mu_r}-\sum\limits_{k=1}^r\partial_\nu X^{\mu_k}T^{\mu_1\ldots\mu_{k-1}\nu\mu_{k+1}\ldots \mu_r}\\
 &=& X^\nu\partial_\nu T^{\mu_1\ldots \mu_r}-r\,\partial_\nu X^{(\mu_1}T^{\mu_2\ldots \mu_r)\nu}\,.
 \label{Lieder}
\end{eqnarray}
The Lie derivative can be written in a suggestive way through the canonical Poisson bracket on the commutative algebra $ C^\infty(T^*M)$ of functions on the cotangent bundle  (as detailed in the next subsection).

\subsubsection{Poisson algebra of symmetric contravariant tensor fields}

A \textbf{Poisson bracket} $\{\,\,,\,\}$ for a commutative algebra
$\mathcal A$ with product $\cdot$ is a Lie bracket which is also a (bi)derivation, \textit{i.e.} $\{x,y\cdot z\}=y\cdot \{x,z\}+\{x,y\}\cdot z$
for any $x,y,z\in{\mathcal A}\,$. A \textbf{Poisson algebra} is both a commutative algebra and a Lie algebra endowed
respectively with an associative product and a Poisson bracket. 
An algebra  $\mathcal A$ is said \textbf{central} if its center coincides with the field $\mathbb K$ of $\mathcal A$.
A Poisson algebra is central as a Lie algebra iff its Poisson bracket is non-degenerate (up to scalars elements).

A \textbf{Poisson manifold} is a manifold whose algebra of functions is a Poisson algebra. 
A \textbf{symplectic manifold} is a Poisson manifold whose algebra of functions is central as a Lie algebra or, equivalently, whose Poisson bracket is non-degenerate (up to constant functions).

The cotangent bundle has a canonical structure of symplectic manifold, whose coordinate-free definition will be introduced later on.
The \textbf{canonical Poisson bracket} is the usual Poisson bracket of Hamiltonian mechanics and reads in coordinates as
\be
\{\,\,,\,\}_{_C}\,:=\,\frac{\overleftarrow{\partial}}{\partial x^{\mu}}\,\frac{\overrightarrow{\partial}}{\partial p_{\mu}}-\frac{\overleftarrow{\partial}}{\partial p_{\mu}}\,\frac{\overrightarrow{\partial}}{\partial x^{\mu}}\,,
\label{Poisson}
\ee
where the arrows indicate on which factor they act. 
Locally, it acts as follows
\be
\{\,P(x,p)\,,Q(x,p)\,\}_{_C}=\frac{\partial P}{\partial x^{\mu}}\frac{\partial Q}{\partial p_{\mu}}-\frac{\partial P}{\partial p_{\mu}}\frac{\partial Q}{\partial x^{\mu}}\,.
\ee

The \textbf{suspension} $V[1]$ of an $\mathbb N$-graded vector space $V$ is the graded vector space whose grading is obtained by shifting down the grading of $V$ by one: $V[1]\,{}_i=V_{i+1}$.
For instance, the suspension ${\mathcal A}[1]$ of an associative algebra is an an $\mathbb N$-graded algebra iff the product $\star$ of the initial algebra ${\mathcal A}$ obeys 
\be
{\mathcal A}_{m+1}\star {\mathcal A}_{n+1}\subseteq{\mathcal A}_{m+n+1}\quad\Leftrightarrow\quad{\mathcal A}_i\star {\mathcal A}_j\subseteq {\mathcal A}_{i+j-1}\,.
\ee
A Poisson algebra ${\mathcal P}$ which is $\mathbb N$-graded as a commutative algebra and such that its suspension ${\mathcal P}[1]$ is $\mathbb N$-graded as a Lie algebra is called a \textbf{Schouten algebra}.\footnote{Note that a Gerstenhaber algebra is the supercommutative analogue of a Schouten algebra.} Concretely, this condition means that the commutative product $\cdot$ of ${\mathcal P}$ is such that ${\mathcal P}_i\cdot{\mathcal P}_j\subseteq {\mathcal P}_{i+j}$ while the Poisson bracket $\{\,,\,\}$ is such that $\{{\mathcal P}_i,{\mathcal P}_j\}\subseteq {\mathcal P}_{i+j-1}$.

\vspace{3mm}
\noindent{\small\textbf{Example (Symbols)\,:} The canonical Poisson bracket endows the space ${\mathcal S}(M)$ of symbols with a structure of Schouten algebra (since the canonical Poisson bracket decreases the grading by one) which is furthermore central.
Due to the isomorphism ${\mathcal S}(M)\cong\Gamma(\odot TM)$ between the vector space of symbols and the vector space of symmetric contravariant tensor fields, one may infer that the latter can be endowed with a structure of Schouten algebra.
}

\vspace{3mm}
\noindent{\small\textbf{Example (Symmetric contravariant tensor fields)\,:}
It can be checked by direct computation that the Lie derivative \eqref{Lieder} of the symmetric contravariant tensor fields encoded into
the function $T(x,p)=
\sum\frac1{r!}\,T^{\mu_1\ldots \mu_r}(x) p_{\mu_1} \cdots p_{\mu_r}$ on the cotangent bundle $T^*M$
merely corresponds to the Poisson bracket with the function $X(x,p)=X^\nu(x)\,p_\nu$ corresponding to the 
vector field $\hat X=X^\mu(x)\partial_\mu$: 
\be
({\mathcal L}_{\hat{X}}T)(x,p)\,=\,\{T(x,p)\,,\,X(x,p)\}_{_C}\,.
\ee
More generally, the canonical Poisson bracket of functions on $T^*M$ with polynomial dependence in the fibre coordinates induces the   \textbf{Schouten bracket of symmetric contravariant tensor fields} (see \textit{e.g.} \cite{DuboisV} and refs therein)
\begin{eqnarray}
\{\,\,,\,\}_S&:&\,\Gamma(\odot^{r_1}TM)\otimes\,\Gamma(\odot^{r_2}TM)
\rightarrow\Gamma(\odot^{r_1+r_2-1}TM)\nonumber\\
&:&T_1^{\,\nu_1\ldots\nu_{r_1}}(x)\,\otimes \,T_2^{\,\nu_1\ldots\nu_{r_2}}(x)\,\,\longmapsto 
\,\,\{\,T_1\,, T_2\,\}_S^{\nu_1\ldots\nu_{r_1+r_2-1}}(x)\,,
\end{eqnarray}
where 
\begin{eqnarray}
\{\, T_1\,, T_2\,\}_S^{\nu_1\ldots\nu_{r_1+r_2-1}}\,\,:=&r_2&\partial_\mu  T_1^{\,(\nu_1\ldots\nu_{r_1}} T_2^{\,\nu_{r_1+1}\ldots\nu_{r_1+r_2-1})\mu}\,\nonumber\\
&&-\,\, r_1\,\, T_1^{\,\mu(\nu_1\ldots\nu_{r_1-1}}\partial_\mu  T_2^{\,\nu_{r_1}\ldots\nu_{r_1+r_2-1})}\,.
\label{Schouten}
\end{eqnarray}
As one can see, together with the symmetric product $\odot$, the Schouten bracket $\{\,\,,\,\}_S$ (which decreases the rank by one) endows the algebra $\Gamma(\odot TM)$ of symmetric contravariant tensors with a structure of Schouten algebra.
}
\vspace{3mm}

A diffeomorphism of a symplectic manifold preserving the Poisson bracket is called a \textbf{symplectomorphism}. Algebraically, a symplectomorphism can be defined as an automorphism of the Poisson algebra of functions on the symplectic manifold, \textit{i.e.} it is an automorphism for both its commutative and its Lie algebra structures. 
A derivation of the Poisson algebra of functions on a symplectic manifold, with respect to \textit{both} the pointwise product and the Poisson bracket, is called a \textbf{symplectic vector field}. A symplectic vector field corresponds to an infinitesimal symplectomorphism since it is a vector field on a symplectic manifold that preserves the Poisson bracket. 

Let $h$ be a function on a symplectic manifold. A \textbf{Hamiltonian vector field} for the function $h$, then called its \textbf{Hamiltonian}, is a symplectic vector field $\hat{X}_h$ whose corresponding Lie derivative is the inner derivation 
\be
{\mathcal L}_{\hat{X}_h}\,=\,\{\,\,,h\}\,,
\ee
\textit{i.e.} $\hat{X}_h[g]=\{g,h\}$ for all functions $g$ on the symplectic manifold. 
A \textbf{symplectic/Hamiltonian flow} is a one-parameter group of symplectomorphisms generated by a symplectic/Hamiltonian vector field.
For instance, a Hamiltonian flow on a symplectic manifold ${\mathcal M}$ is a one-parameter group of symplectomorphisms $\exp(t\hat{X}_h)$ of ${\mathcal M}$ generated by a Hamiltonian vector field $\hat{X}_h\in\mathcal{T}({\mathcal M})$ defined by a Hamiltonian $h\in C^\infty({\mathcal M})$.
All symplectic flows are locally Hamiltonian because, locally, all symplectic vector fields (\textit{i.e.} derivations of the Poisson algebra of functions on a symplectic manifold) are Hamiltonian vector fields (\textit{i.e.} inner derivations).\footnote{To be more precise, the number of linearly independent outer derivations of the Poisson algebra of functions on ${\mathcal M}$ is equal to the first Betti number of ${\mathcal M}$ (which vanishes for a topologically trivial manifold $M$).} This remains true for the Schouten algebra of symbols on $M$ \cite{Grabowski:2003b}: locally, all derivations of the Schouten algebra ${\mathcal S}(M)$ are inner, \textit{i.e.} Hamiltonian vector fields $\hat{X}_h$ with $h\in{\mathcal S}(M)$. 

\pagebreak

\section{Differential structures of order one: first-order (co)jets}\label{diffopss}

In quantum mechanics (and, more generally, in classical or quantum field theory), infinitesimal space-time symmetries must sometimes be represented as differential operators of order one rather than vector fields (this happens, for instance, when one considers projective representations). Therefore, field theory is a physical motivation for considering 
first-order differential operators as a natural generalisation of vector fields, discussed in this section.

\subsection{Functions as differential operators of order zero}

Any function $f$ on a manifold $M$ can also be seen as a linear operator $\hat{f}$ on the structure algebra $ C^\infty(M)$ acting by multiplication, \textit{i.e.}
mapping any function $g$ on $M$ to the function
\be
\hat{f}[\,g]\,:=\,f\cdot g\,,
\ee
where $\cdot$ is the pointwise product.
Retrospectively, the function $f$ can be reconstructed from the corresponding linear operator $\hat{f}$ via the action of the operator on the constant function equal to unity, $\hat{f}[1]=f\cdot 1=f\,$.

The structure algebra $C^\infty(M)$ of the manifold $M$ is a $ C^\infty(M)$-(bi)module via the pointwise product.
A $ C^\infty(M)$-linear operator on $C^\infty(M)$ is called a \textbf{differential operator of order zero} on $M$.
The space of differential operators of order zero will be denoted by ${\mathcal D}^0(M)$.

Due to the above remark, any function $f$ on $M$ can be seen as a differential operator $\hat{f}$ of order zero on $M$, and conversely. In other words, ${\mathcal D}^0(M)\cong C^\infty(M)$. For the sake of simplicity, from now on differential operators of order zero will sometimes be identified with functions.

\subsection{First-order differential operators}

A \textbf{differential operator of order one} on $M$ can be defined as a linear operator $\hat{X}$ on $ C^\infty(M)$ 
such that its commutator with any differential operator $\hat{f}$ of order zero is also a differential operator  of order zero:
$[\hat{X},\hat{f}]\,
\in {\mathcal D}^0(M)$.
The space of differential operators on $M$ of order one will be denoted by ${\mathcal D}^1(M)$. The commutator closes on first-order differential operators.
Therefore, the space of first-order differential operators  ${\mathcal D}^1(M)$ can be endowed with a 
structure of Lie algebra with the commutator as Lie bracket. In such case, it will sometimes be denoted ${\mathfrak{D}}^1(M)$ in order to emphasise its Lie algebra structure.

\subsubsection{Vector fields as first-order differential operators}

An equivalent definition of a vector field on $M$ is as 
a differential operator $\hat{X}$ of order one such that its action on any constant function on $M$ is zero, 
\textit{e.g.} $\hat{X}[1]=0$.

\vspace{3mm}

\noindent{\small\textbf{Proof:} One can check that if an operator $\hat{X}$ is such that 
$\hat{X}[1]=0$ and, moreover, $[\hat{X},\hat{f}]$ is a zeroth-order operator for any function $f$, then
its action on a function $f$ (that is to say $\hat{X}[f]$)
identifies with its adjoint action (in other words $[\hat{X},\hat{f}]$) on the associated operator $\hat{f}$. Indeed,
\be
\hat{X}[f]=\hat{X}[f\cdot 1]=\hat{X}\big[\,\hat{f}[1]\,\big]
=(\hat{X}\circ\hat{f})[1]=(\hat{X}\circ\hat{f}-\hat{f}\circ\hat{X})[1]=\left(\,[\hat{X},\hat{f}]\right)[1]\,.
\ee
In a similar way, one can then obtain the property that $\hat{X}$ is a derivation as a consequence of
the Jacobi identity for the commutator.\qed
}

\vspace{3mm}

The previous definitions lead to the following unique (and coordinate-independent) decomposition of any first order differential operator into a sum $\hat{X}=\hat{X}_0+\hat{X}_1$ of a zeroth-order differential operator $\hat{X}_0$, associated to the function $X_0:=\hat{X}[1]$, and a vector field $\hat{X}_1:=\hat{X}-\hat{X}_0$. 

\subsubsection{Semidirect sum of Lie algebras}\label{semidriectsums}

A \textbf{semidirect sum of Lie algebras} is a Lie algebra decomposing as the linear sum of a Lie subalgebra $\mathfrak{h}\subset\mathfrak{g}$ and a Lie ideal $\mathfrak{i}\subset\mathfrak{g}$. It is sometimes denoted in the following ways:
\be
\mathfrak{g}\,=\,\mathfrak{h}\inplus\mathfrak{i}\,=\,\mathfrak{i}\niplus\mathfrak{h}\,.
\ee
Concretely, the Lie bracket obeys 
\be
[\mathfrak{h},\mathfrak{h}]\subset\mathfrak{h}\,,\quad[\mathfrak{i},\mathfrak{i}]\subset\mathfrak{i}\,,\quad[\mathfrak{h},\mathfrak{i}]\subset\mathfrak{i}\,.
\ee
Therefore the linear decomposition $\mathfrak{g}=\mathfrak{h}\inplus\mathfrak{i}$ is actually a direct sum of $\mathfrak{h}$-modules.  

In fact, one can define equivalently a semidirect sum as a Lie algebra $\mathfrak{g}$ with a Lie subalgebra $\mathfrak{h}\subset\mathfrak{g}$ such that the adjoint representation $ad^\mathfrak{g}:\mathfrak{g}\to\mathfrak{der}(\mathfrak{g})$
of the Lie algebra $\mathfrak{g}$ on itself, restricted to the adjoint representation $ad^\mathfrak{g}|_{\mathfrak{h}}:\mathfrak{h}\to\mathfrak{der}(\mathfrak{g})$ of the Lie subalgebra $\mathfrak{h}\subset\mathfrak{g}$ on the whole Lie algebra $\mathfrak{g}$ is fully reducible. It decomposes as the direct sum $ad^\mathfrak{g}|_{\mathfrak{h}}=ad^\mathfrak{h}\oplus r$ of 
\begin{enumerate}
  \item the adjoint representation $ad^\mathfrak{h}:\mathfrak{h}\to\mathfrak{der}(\mathfrak{h})$ of the Lie subalgebra $\mathfrak{h}$ on itself, and
	\item a representation
\be\label{rephoni}
r\,:\,\mathfrak{h}\to\mathfrak{der}(\mathfrak{i})\,:\,v\mapsto r_v
\ee
of the Lie subalgebra $\mathfrak{h}$ on the Lie ideal $\mathfrak{i}$ with $r_v(w):=[v,w]$ for $v\in\mathfrak{h}$ and $w\in\mathfrak{i}$.
\end{enumerate} 
Conversely, given two Lie algebras $\mathfrak{h}$ and $\mathfrak{i}$ together with a representation \eqref{rephoni} of the Lie algebra $\mathfrak{h}$ on the Lie algebra $\mathfrak{i}$ one can form their semidirect sum $\mathfrak{g}=\mathfrak{h}\inplus_r\mathfrak{i}$ 
by endowing their linear sum with a Lie algebra structure via the Lie bracket
\be
[\,v_1\oplus w_1\,,\,v_2\oplus w_2\,]_\mathfrak{g}=[\,v_1\,,\,v_2\,]_\mathfrak{h}\,\oplus\,\Big(\,r_{v_1}(w_2)-r_{v_2}(w_1)+[\,w_1\,,\,w_2\,]_\mathfrak{i}\,\Big)
\ee
for all $v_1,v_2\in\mathfrak{h}$ and $w_1,w_2\in\mathfrak{i}$.

\vspace{3mm}
\noindent{\small\textbf{Example (First-order differential operators)\,:} For instance, the Lie algebra of first-order differential operators splits into the semidirect sum 
\be\label{semidir1storder}
{\mathfrak{D}}^1(M)\cong\mathfrak{X}(M)\inplus  C^\infty(M)
\ee
of the Lie subalgebra $\mathfrak{X}(M)$ of vector fields and the Abelian ideal ${\mathfrak{D}}^0(M)\cong C^\infty(M)$ of differential operators of order zero.
}

\subsubsection{Some general facts}

A theorem of Grabowski and Poncin provides an analogue of the generalised Milnor exercise for the Lie algebra of first-order differential operators \cite{Grabowski:2002}.

\vspace{3mm}
\noindent{\textbf{Theorem (Grabowski \& Poncin)\,:}
\textit{A map $\Phi:{\mathfrak{D}}^1(M)\stackrel{\sim}{\to}{\mathfrak{D}}^1(N)$ between the Lie algebras of first-order differential operators on two manifolds is an isomorphism of Lie algebras iff there is a diffeomorphism $F:M\stackrel{\sim}{\to} N$ between these two manifolds.}}
\vspace{3mm}

On the one hand, the Lie algebra $\mathfrak{X}(M)$ of vector fields may be endowed with a structure of left ${\mathcal C}^\infty(M)$-module via the composition product $\circ$ (which will often be implicit in the sequel)
since the operator $\hat{f}\circ\hat{X}$ is again a vector field if $f$ is a function and $\hat X$ is a vector field.
The previous product will be called \textbf{pointwise product} of a function $f$ and a vector field $\hat{X}$, and it will be denoted as $f\cdot \hat{X}:=\hat{f}\circ\hat{X}\,$.
However, the Lie bracket of vector fields is not ${\mathcal C}^\infty(M)$-linear, rather it obeys to the Leibniz rule
\be
 [\,\hat{X}\,,\,f\cdot\hat{Y}\,]\,=\,\hat{X}[f]\cdot\hat{Y}\,+\,f\cdot[\hat{X},\hat{Y}]\,\,,
\label{Leibnizz}
\ee
for $\mathfrak{X}(M)$ seen as a left ${\mathcal C}^\infty(M)$-module.\footnote{By default, the modules will be considered to be left modules
therefore the word ``left'' will often be implicit.}

On the other hand, the space ${\mathcal D}^1(M)$ of first-order differential operators 
is a left and right $C^\infty(M)$-module, \textit{i.e.}
a $C^\infty(M)$-bimodule,
since  $\hat{f}\circ\hat{X}\circ\hat{g}$ is a first-order differential operator if $f$ and $g$ are functions while $\hat X$ is a first-order differential operator.
Notice that this composition product $\hat{X}\circ\,\hat{g}$ with a function on the right is no more pointwise (\textit{i.e.} it does not only depend on the values of the objects at the same point) since it includes a derivative of the function $f$.
More explicitly, for a first order differential operator decomposing into the sum $\hat{X}=\hat{X}_1+\hat{X}_0$ of a vector field $\hat{X}_1$ and a zeroth-order differential operator $\hat{X}_0$ one has
\be
\hat{X}\circ\hat{f}\,=\,[\hat{X}\stackrel{\circ}{,}\hat{f}]\,+\,\hat{f}\circ\hat{X}\,=\,\hat{X}_1[f]\cdot\,+\,f\cdot\hat{X}\,.
\ee

\subsection{First-order jets}

Let $m$ be a point of $M$. Consider the commutative ideal ${\mathcal I}^1(m)$ of $ C^\infty(M)$
spanned by the functions $f$ such that $f|_m=0$ and $df|_m=0$, \textit{i.e.} the functions whose value and first derivatives vanish at the point $m$. It will be called \textbf{contact ideal of order one}.
The quotient of the contact ideal of order zero by the contact ideal of order one is isomorphic to the cotangent space: 
\be\label{T*MI0I1}
T^*_mM\,\cong\,{\mathcal I}^0(m)\,/\,{\mathcal I}^1(m)\,,
\ee
as can be expected since one retains only the information about the first derivative of functions at $m$ and not about the value of the function (at the same point $m$).

The contact ideal of order one ${\mathcal I}^1(m)\subset C^\infty(M)$ defines an equivalence relation whose equivalence classes $[f]$,  denoted by $j^1_m f$, are called \textbf{jets of order one} or \textbf{first-order jets} (or simply \textbf{1-jets}) of functions.
More concretely, two functions $f$ and $g$ are said to have the same jet $j^1_m f$ of order one at $m$ 
if they only differ by an element of the commutative ideal ${\mathcal I}^1(m)$, 
\textit{i.e.} if these two functions together with their first derivatives have the same
value at $m$.
The point $m$ is sometimes called the source of the jet $j^1_m f$.
The quotient 
\be
J^1_mM\,:=\,  C^\infty(M)\,/\,{\mathcal I}^1(m)
\ee
is called the \textbf{first-order jet space} at $m$.
From the isomorphism \eqref{T*MI0I1}, it follows that the cotangent space is isomorphic to the quotient of the first-order jet space by the zeroth-order one: 
\be
T_m^*M\,\cong\,J^1_mM\,/\,J^0_mM\,.
\ee

The \textbf{first-order jet bundle} is the manifold $J^1M:=\bigcup_m J^1_m M $ and it has local coordinates $(x^\mu,\phi,\phi_\nu)$.
As one may suspect from the coordinate expression and the above isomorphisms, the first-order jet bundle is isomorphic to the direct sum over $M$ of the zeroth-order jet bundle $J^0M=M\times {\mathbb R}$ and the cotangent bundle $T^*M$, \textit{i.e.} 
\be
J^1M\cong J^0M\oplus T^*M\,.
\ee  
The first-order jet bundle is the paradigmatic example of ``contact manifold'' but one will not dwelve on this notion.

A compact notation for a 1-jet is as a polynomial of degree one in an auxiliary variables $\varepsilon^\mu$, \textit{i.e.} $\phi(\varepsilon)=\phi+\varepsilon^\mu\phi_\mu$.
A section of the 1-jet bundle $J^1M$ is called a \textbf{$1$-jet field} therefore compactly summarised into the local expression \be
\phi(x;\varepsilon)=\phi(x)+\varepsilon^\mu\phi_\mu(x)\,.
\ee
The commutative algebra of $1$-jet fields will be denoted $\mathcal{J}^1(M):=\Gamma(J^1M)$.

The \textbf{first-order prolongation of a function} $f\in  C^\infty(M)$ is a $1$-jet field $j^1f\in\mathcal{J}^1(M)$ whose value at a point $m$ is the $1$-jet $j^1_mf$ of the function. Its local expression is 
\be\label{j1f}
(j^1 f)(x;\varepsilon)=f(x)+\varepsilon^\mu \partial_\mu f(x),
\ee
\textit{i.e.} $(j^1 f)(x;0)=f(x)$ and $(j^1 f)_\mu(x)=\partial_\mu f(x)$.
This defines the \textbf{first-order prolongation} $j^1: C^\infty(M)\to \mathcal{J}^1(M)$.

\vspace{1mm}
{\small In more geometrical terms, the $1$-jet $j_m^1f$ at a point $m$ can be interpreted as the equivalence classes of sections of the structure bundle that touch each other till order $1$. In other words, they are (nowhere vertical) codimension-one submanifolds that intersect and are tangent at the point $(\,m\,,f|_m\,)\in M\times {\mathbb R}$.}

\subsection{First-order cojets}

The dual space $D^1_mM:=(J^1_mM)^*$ of linear forms on the first-order jet space $J^1_mM$ is called the \textbf{1-cojet space} at $m$. 
In local coordinates, a compact notation for a \textbf{1-cojet} (or \textbf{first-order cojet}) $X$ is as a polynomial of degree one in the auxiliary variables $\frac{\partial}{\partial \varepsilon^\mu}$, \textit{i.e.} 
$X(\,\partial_\varepsilon)=X^\mu \frac{\partial}{\partial \varepsilon^\mu} +X$.
Concretely, in this representation a 1-cojet $X$ acts on a 1-jet $\phi$ at the same point, as a first-order differential operator with respect to the auxiliary coordinate $\varepsilon$ followed by an evaluation at $\varepsilon=0$:
\be
\langle X, \phi\,\rangle =\left.\big(\, X(\partial_\varepsilon)\,\phi(\varepsilon)\,\big)\,\right|_{\varepsilon=0} = X^\mu \phi_\mu +X\phi\,.
\label{duality1}
\ee
since the representative of $\phi$ is the polynomial $\phi(\,\varepsilon)=\phi+\varepsilon^\mu\phi_\mu$ of degree one.

The \textbf{1-cojet bundle} $D^1M=\bigcup_m D^1_mM$ has local coordinates $(x^\mu,X,X^\nu)$.
A section of the 1-cojet bundle $D^1M$ will be called a \textbf{1-cojet field} and is compactly summarised into
a generating function $X(x;\partial_\varepsilon)=X^\mu(x) \frac{\partial}{\partial \varepsilon^\mu} +X(x)$.
The space $Gamma(D^1M)$ of such sections  will be denoted by the same symbol ${\mathcal D}^1(M)$ as the space of differential operators of order one since they are isomorphic. Notice that any 1-cojet field can be interpreted as a map $X:\mathcal{J}^1(M)\to  C^\infty(M)$ where the explicit action of a 1-cojet field $X(x,\partial_\varepsilon)$ in ${\mathcal D}^1(M)$ 
on a 1-jet field $\phi(x;\varepsilon)$ in $\mathcal{J}^1(M)$ is given by \eqref{duality1},
where the dependence on $x$ would be implicit.

Retrospectively, first-order differential operators on $M$ can be defined as linear operators
\be
\hat{X}\,:\, C^\infty(M)\to  C^\infty(M)\,:\,f\mapsto \hat X[f]
\ee which factor through the 1-jet bundle 
$J^1M$ in the sense that 
\be
\hat{X}={X}\circ j^1\,,
\ee
 where ${X}:\mathcal{J}^1(M)\to  C^\infty(M)$ is a 1-cojet field in ${\mathcal D}^{1}(M)$
and $j^1: C^\infty(M)\to \mathcal{J}^1(M)$ is the first-order prolongation.
Concretely this means that a first-order differential operator whose local expression is $\hat{X}=X^\mu(x)\frac{\partial}{\partial x^\mu} + X(x)$ obviously defines a 1-cojet field ${X}(x;\partial_\varepsilon)=X^\mu(x) \frac{\partial}{\partial \varepsilon^\mu} +X(x)$. Moreover, the action of  $\hat{X}={X}\circ j^1$ on $f$ is indeed
\begin{eqnarray}
(\hat{X}f)(x)&=& X^\mu(x)\,\partial_\mu f(x)\, +\, X(x)\,f(x)\nonumber\\
&=&\left.\big(\, X(x;\partial_\varepsilon)\,(j^1f)(x;\varepsilon)\,\big)\,\right|_{\varepsilon=0}\nonumber\\
&=&\langle \,X\,,\,j^1f\,\rangle (x)\,. 
\end{eqnarray}
where one made use of \eqref{j1f} to obtain the second line.

\pagebreak

\section{Differential structures of higher order: (co)jets}\label{co/jets}

The time is ripe to generalise the previous constructions and aim at our goal: provide a global and coordinate-free definition of higher-order differential operators (as well as their relatives, such as cojet fields, etc).

\subsection{Higher-order differential operators}

\subsubsection{Rings over an algebra}

Let $\mathcal A$ and $\cal B$ be two associative algebras with respective zeros $0_{\mathcal A}$ and $0_{\mathcal B}$, and respective units $1_{\mathcal A}$ and $1_{\mathcal B}$, and respective products $\star_{\mathcal A}$ and $\star_{\mathcal B}$. 

An injective morphism $i:{\mathcal A}\hookrightarrow{\mathcal B}$ of algebras,
\be
i(a_1\star_{\mathcal A} a_2)=i(a_1)\star_{\mathcal B} i(a_2)\,,\qquad\forall a_1,a_2\in{\mathcal A}\,,
\ee 
will be called a \textbf{unit map}.\footnote{This terminology is standard in the context of bialgebroids, where the assumption of injectivity is dropped. The terminology originates from the fact that, in particular, the unit map somehow relates the two units in the sense that $i(1_{\mathcal A})=1_{\mathcal B}$). Note that the injectivity can be assumed without loss of generality, in the sense that one can always focus on the quotient algebra ${\mathcal A}/\ker\,i$.} An associative algebra $\cal B$ endowed with a unit map $i:{\mathcal A}\hookrightarrow{\mathcal B}$ will be called a \textbf{ring} $\cal B$ \textbf{over the algebra} $\mathcal A$ (or an $\mathcal A$-\textbf{ring} for short). Equivalently, $\cal B$ admits a subalgebra isomorphic to $\mathcal A$: the image $i({\mathcal A})\subseteq\cal B$. The unit map endows an $\mathcal A$-ring $\cal B$ with a structure of $\mathcal A$-bimodule where an element $a\in\mathcal A$ acts by (left or right) multiplication by $i(a)\in\cal B$. Conversely, an $\mathcal A$-bimodule structure on $\cal B$ such that $a\bullet 1_{\mathcal B}=0_{\mathcal B}$ iff $a=0_{\mathcal A}$, endows the associative algebra $\cal B$ with a structure of $\mathcal A$-ring, by considering the action of $\mathcal A$ on the unit element $1_{\mathcal B}$ (\textit{i.e.} defining the unit map by $i(a):=a\bullet 1_{\mathcal B}$ for all $a\in\mathcal A$).

\vspace{5mm}\begin{figure}[h!]
\begin{framed}
\begin{center}
\textbf{Equivalent formulations of $\mathcal A$-rings}
\end{center}

\noindent
Given two associative algebras $\mathcal A$ and $\cal B$, the following notions are equivalent:
\begin{enumerate}
 \item an $\mathcal A$-ring $\cal B$ defined by an injective algebra morphism $i:{\mathcal A}\hookrightarrow{\mathcal B}$.
 \item an algebra $\cal B$ containg a copy of $\mathcal A$, \textit{i.e.} the image $i({\mathcal A})\subseteq\cal B$.
 \item an $\mathcal A$-bimodule structure on $\cal B$ such that the action of the nonvanishing elements of ${\mathcal A}$ on the unit element of ${\mathcal B}$ is nonvanishing.
\end{enumerate}
\vspace{3mm}
\end{framed}
\end{figure}

\vspace{3mm}
\noindent{\small\textbf{Example (Endomorphism algebra)\,:} Let ${\mathcal A}$ be an associative algebra whose product $\cdot$ is non-degenerate (in the sense that $f\cdot g=0$ for all $g\in\mathcal A$ iff $f=0$). Any element $f$ of ${\mathcal A}$ defines an ${\mathcal A}$-linear map $\hat{f}\in\text{End}({\mathcal A})$ acting on ${\mathcal A}$ by multiplication by $f$, \textit{i.e.} in a compact notation $\hat{f}:=f\cdot$ or, more explicitly, $\hat{f}:g\mapsto f\cdot g$. The left multiplication provides a canonical representation 
\be\label{hatbull}
\hat{\bullet}\,:\,{\mathcal A}\hookrightarrow \text{End}({\mathcal A})\,:\,f\mapsto\hat{f}\,,
\ee
of $\cal A$ on itself which is faithful (since the product is non-degenerate, a property which will always be assumed implictly below).
}

\subsubsection{Algebraic definition of differential operators}

Let $\mathcal B$ be an associative algebra with a commutative subalgebra ${\mathcal A}\subset{\mathcal B}$. An element $a\in{\mathcal B}$ such that, for any set $\{f_1, f_2, \cdots,f_k\}\subset{\mathcal A}$ of $k$ elements in the commutative subalgebra ${\mathcal A}$, the $k$th commutator with each of them belongs to ${\mathcal A}$, \textit{i.e.}
\be
[\,[\,\ldots\,[a\,,\,f_1]\,,\,f_2]\ldots \,,\,f_k]\in {\mathcal A}\,,
\ee
will be called a \textbf{differential element of order} $k$ \textbf{with respect to the commutative subalgebra} ${\mathcal A}$. The motivation for this terminology is the particular case of scalar-valued linear operators on a commutative algebra ${\mathcal A}$.

A (scalar-valued linear) \textbf{differential operator of order} $k$ \textbf{acting on the commutative algebra} ${\mathcal A}$
is an endomorphism $\hat{X}\in\text{End}({\mathcal A})$ such that, for any set $\{f_1, f_2, \cdots,f_k\}\subset{\mathcal A}$ of $k$ elements in the commutative algebra ${\mathcal A}$ (identified with the commutative algebra of zeroth-order differential operators on ${\mathcal A}$),
the $k$th commutator with each of them belongs to ${\mathcal A}$, in the sense that
\be\label{kthcommutator}
[\,[\,\ldots\,[\hat{X}\,,\,\hat{f}_1]\,,\,\hat{f}_2]\ldots \,,\,\hat{f}_k]\in{\mathcal D}^0({\mathcal A})\,.
\ee
This abstract definition of differential operators is by now standard and was introduced in the sixties by Grothendieck in his seminal \textit{\'El\'ements de g\'eom\'etrie alg\'ebrique} \cite[Section 16.8]{Grothendieck:1967}.

This provides a purely algebraic (and coordinate-free) definition of a \textbf{differential operator of order} $k$ \textbf{on the manifold} $M$ as a (scalar-valued) linear differential operator of order $k$ acting on the structure algebra $C^\infty(M)$ of functions on $M$.
A convenient representation of a differential operator $\hat{X}$ of order $k$ is through its \textbf{normal symbol in some local coordinates} 
\be
X_{\text{normal}}(x,p)=\sum\limits_{r=0}^k \frac1{r!}\,X^{\mu_1\ldots \mu_r}(x)\,p_{\mu_1}\ldots p_{\mu_r}
\label{standard}
\ee 
which is a symbol of degree $k$ corresponding
to the ``normal ordering'' of the operator
\be
\hat{X}(x,\partial) =\sum\limits_{r=0}^k \frac1{r!}\,X^{\mu_1\ldots \mu_r}(x)\,\partial_{\mu_1}\ldots \partial_{\mu_r}\,,
\label{xdiffop}
\ee
where all derivatives are on the right and all coordinates on the left.\footnote{Strictly speaking, this is not the normal ordering prescription in quantum mechanics if $x$'s are treated as position operators. Rather, here $x$'s are treated as creation operators while $\partial$'s are treated as annihilation operators. Despite the inaccurate terminology, we will use the adjective ``normal'' (order, symbol, etc) in this case, following the common usage in deformation quantisation.} Note that this choice of ordering is not a coordinate-independent statement.

The multi-index notation consists in denoting the collection $\mu_1\ldots\mu_r$ of $r$ symmetrised indices as $\mu(r)$. Moreover, there will be an implicit sum over repeated multi-indices and over the number $r$ of them. More precisely, one will stick to the weight one convention, \textit{i.e.} $$S^{\mu(r)} T_{\mu(r)}:=\sum\limits_r\frac1{r!}\,S^{\mu_1\ldots \mu_r}\,T_{\mu_1\ldots \mu_r}\,.$$
Adopting the multi-index convention, one could have written \eqref{xdiffop} simply as a suggestive generalisation of vector fields
$\hat{X} = X^{\mu(r)}(x)\,\partial_{\mu(r)}$
by introducing the convenient notation 
$\partial_{\mu(r)}:=\partial_{\mu_1}\ldots \partial_{\mu_r}$\,.
The normal symbol \eqref{standard} is obtained from the action of the operator \eqref{xdiffop} on the exponential function (which would be a ``plane wave'' in quantum mechanics), \textit{i.e.} 
\be
X_{\text{normal}}(x,p)\,:=\,\exp(-p_\mu x^\mu)\,\hat{X}[\,\exp(p_\mu x^\mu)\,]\,.
\ee  
The \textbf{principal symbol of a differential operator} \eqref{xdiffop} is its leading (highest order) part 
\be\label{principsymb}
X_{\text{principal}}(x,p)=\frac1{r!}\,X^{\mu_1\ldots \mu_k}(x)\,p_{\mu_1}\ldots p_{\mu_k}\,,
\ee
which admits a coordinate-independent definition.

The space of differential operator of order $k$ will be denoted by ${\mathcal D}^k(M)$, while the space of
all differential operators (of any finite order) will be denoted by ${\mathcal D}(M)$. 
The collection $\{\partial_{\mu_1}\ldots \partial_{\mu_r}\,|\,r\leqslant k\}$ of commuting differential operators is a finite generating set of the $ C^\infty(M)$-module ${\mathcal D}^k(M)$ which will be called the \textbf{coordinate basis}.

\vspace{3mm}
\noindent{\small\textbf{Example (Grothendieck algebra)\,:} The algebra morphism
\be\label{hatbullettt}
\hat{\bullet}\,:\,{\mathcal A}\hookrightarrow{\mathcal D}({\mathcal A})\,:\,f\mapsto\hat{f}\,,
\ee
which reinterprets elements $f$ of a commutative algebra $\mathcal A$ as zeroth-order differential operators $\hat{f}$ on $\mathcal A$, via left multiplication, endows the Grothendieck algebra ${\mathcal D}({\mathcal A})$ of differential operators on $\mathcal A$ with the structure of $\mathcal A$-ring.
The corestriction $\hat{\bullet}:{\mathcal A}\stackrel{\sim}{\to}{\mathcal D}^0({\mathcal A})$ provides an isomorphism between the commutative algebra $\mathcal A$ and the subalgebra ${\mathcal D}^0({\mathcal A})$ of zeroth-order differential operators on $\mathcal A$.}

\vspace{3mm}
\noindent{\small\textbf{Example (Differential operators on a manifold)\,:} 
The canonical embedding of commutative algebras from the structure algebra $ C^\infty(M)$ into the algebra ${\mathcal D}(M)$ of differential operators on the manifold $M$, which reinterprets functions as zeroth-order differential operators,
\be\label{hatbullett}
\hat{\bullet}\,:\, C^\infty(M)\hookrightarrow{\mathcal D}(M)\,:\,f\mapsto\hat{f}\,,
\ee
will be called the \textbf{unit map on the algebra of differential operators on the manifold} $M$.
It induces a canonical isomorphism of commutative algebras from $ C^\infty(M)$ to ${\mathcal D}^0(M)$,
\be\label{hatbullet}
\hat{\bullet}\,:\, C^\infty(M)\stackrel{\sim}{\to}{\mathcal D}^0(M)\,:\,f\mapsto\hat{f}\,.
\ee
}

\subsubsection{Filtration}

An $\mathbb N$-\textbf{filtered} vector space $V$ is a collection $\{V_i\}$ of vector spaces where ${i\in\mathbb N}$ and such that $V_i\subset V_j$ for $i<j$ and $V=\bigcup_{i\in\mathbb N} V_i$. 
An algebra $\mathcal A$ is $\mathbb N$-filtered if it is 
so as a vector space and 
if its product $\star$ obeys ${\mathcal A}_i\star {\mathcal A}_j\subseteq {\mathcal A}_{i+j}$. Moreover, if the algebra $\mathcal A$ has a unit element, then one further requires that $1\in {\mathcal A}_0$.
The component ${\mathcal A}_0$ of degree zero of any $\mathbb N$-filtered algebra $\mathcal A$ is always a subalgebra. In other words,
the $\mathbb N$-filtered algebra $\mathcal A$ is a ring over ${\mathcal A}_0$ (in the absence of left divisors of zero).

The \textbf{filtered algebra associated to a graded algebra} ${\mathcal B}=\oplus_{i\in\mathbb N} {\mathcal B}_i$ is the filtered algebra which will be denoted ${\mathcal B}_{\leqslant}=\bigcup_{i\in\mathbb N} {\mathcal B}_{\leqslant i}$ and which is defined via the direct sums $ {\mathcal B}_{\leqslant i}=\oplus_{j\leqslant i}\, {\mathcal B}_j$\,.
Conversely, the \textbf{graded algebra associated to a filtered algebra} ${\mathcal A}=\bigcup_{i\in\mathbb N} {\mathcal A}_i$ is denoted $\text{gr}{\mathcal A}=\oplus_{i\in\mathbb N}\,\text{gr}{\mathcal A}_i$ and defined via the quotients $\text{gr}{\mathcal A}_i={\mathcal A}_i/{\mathcal A}_{i-1}$. 
The equivalence class $[a]\in \text{gr}{\mathcal A}$ of an element $a\in{\mathcal A}$ of a filtered algebra is an element of the associated graded algebra called the \textbf{principal symbol} of $a$. 
This defines an infinite collection of surjective linear maps
\be\label{principsymbi}
\sigma_i\,:\,{\mathcal A}_i\twoheadrightarrow\text{gr}{\mathcal A}_i\,:\,a\mapsto[a]\,,
\ee 
which will be collectively denoted (with a slight abuse of notation) as
\be\label{principsymb}
\sigma\,:\,{\mathcal A}\twoheadrightarrow\text{gr}{\mathcal A}\,:\,a\mapsto[a]\,.
\ee 

\vspace{3mm}
\noindent{\small\textbf{Example (Polynomials)\,:} For instance, the algebra $\odot {\mathbb R}^{n*}$ of polynomial functions on the affine (respectively, vector) space ${\mathbb R}^{n}$ is the filtered (respectively, graded) algebra of polynomials of $n$ variables. The distinction between these two cases comes from the fact that translations do not preserve the grading while they preserve the filtration. The principal symbol of a polynomial of degree $k$ is identified with its homogeneous piece of highest degree $k$.
}

\subsubsection{Almost-commutative algebra of differential operators}

The $\mathbb N$-graded algebra $\text{gr}{\mathcal A}$ associated to an $\mathbb N$-filtered associative algebra $\mathcal A$ is commutative iff the commutator obeys to $[{\mathcal A}_i,{\mathcal A}_j]\subseteq {\mathcal A}_{i+j-1}$. In such case, the $\mathbb N$-filtered associative algebra $\mathcal A$ is called \textbf{almost commutative}. 

Equivalently, an almost-commutative algebra $\mathcal A$ is an $\mathbb N$-filtered (\textit{i.e.} ${\mathcal A}_i\,{\mathcal A}_j\subseteq {\mathcal A}_{i+j}$) associative algebra whose filtration is such that the suspension $\mathfrak{A}[1]$ of its commutator algebra $\mathfrak{A}$ is a filtered Lie algebra (\textit{i.e.} $[{\mathfrak{A}}_{i+1},{\mathfrak{A}}_{j+1}]\subseteq {\mathfrak{A}}_{i+j+1}$). This equivalent definition makes clear that the $\mathbb N$-graded algebra $\text{gr}{\mathcal A}$ of an almost-commutative algebra $\mathcal A$ is endowed with a canonical structure of Schouten algebra, where the Poisson bracket $\{\,\,,\,\}$ is inherited from the commutator bracket $[\,\,\stackrel{\star}{,}\,]$ via the principal symbol:
\be
\{\sigma(a),\sigma(b)\}:=\sigma\Big(\,[\,a\,\stackrel{\star}{,}\,b\,]\,\Big)\,.
\ee
The Schouten algebra $\text{gr}{\mathcal A}$ associated to an almost commutative algebra $\mathcal A$ will be called the \textbf{classical limit of the almost-commutative algebra}. 
Accordingly, some authors use the term ``quantum'' (respectively ``classical'') Poisson algebra $\mathcal A$ as a synonym for ``almost-commutative'' (respectively, ``Schouten'') algebra \cite{Grabowski:2002}.
Note that an almost-commutative algebra is central iff its classical limit is central. 

Another equivalent way to characterise an almost-commutative algebra is as an associative algebra $\mathcal A$ with a commutative subalgebra ${\mathcal A}_0\subset{\mathcal A}$ such that all elements of $\mathcal A$ are differential with respect to ${\mathcal A}_0$.
Indeed, such an algebra $\mathcal A$ is filtered by the order of elements and one can check that this filtration obeys to $[{\mathcal A}_i,{\mathcal A}_j]\subseteq {\mathcal A}_{i+j-1}$.

\vspace{5mm}\begin{figure}[h!]
\begin{framed}\noindent
\begin{center}
\textbf{Equivalent formulations of almost-commutative algebras}
\end{center}

\noindent
The following notions are equivalent:
\begin{enumerate}

\item an almost-commutative algebra,

\item an $\mathbb N$-filtered associative algebra $\mathcal A$ such that one of the following equivalent properties holds:

	\begin{enumerate}

	\item its associated  $\mathbb N$-graded algebra $\text{gr}{\mathcal A}$ is commutative,

	\item its commutator decreases the degree by one: $[{\mathcal A}_i,{\mathcal A}_j]\subseteq {\mathcal A}_{i+j-1}$,

	\item the suspension $\mathfrak{A}[1]$ of its commutator algebra $\mathfrak{A}$ is a filtered Lie algebra: $[{\mathfrak{A}}_{i+1},{\mathfrak{A}}_{j+1}]\subseteq {\mathfrak{A}}_{i+j+1}$.
\end{enumerate}

\item an associative algebra $\mathcal A$ with a commutative subalgebra ${\mathcal A}_0\subset{\mathcal A}$ such that all elements of $\mathcal A$ are differential with respect to ${\mathcal A}_0$.
\end{enumerate}
\vspace{3mm}
\end{framed}
\end{figure}
\vspace{5mm}
 
\noindent{\small\textbf{Example (Grothendieck algebra)\,:} The composition product $\circ$ endows the space ${\mathcal D}({\mathcal A}_0)$ of
all differential operators acting on a commutative algebra ${\mathcal A}_0$ with a structure of almost-commutative algebra. This algebra will be called the \textbf{Grothendieck algebra of differential operators acting on the commutative algebra} ${\mathcal A}_0$.}

\vspace{3mm}
These abstract constructions can be illustrated in the two simpler cases of polynomial and formal power series, before turning back to smooth functions.

\vspace{3mm}
\noindent{\small\textbf{Example (Polynomial differential operators)\,:} The Grothendieck algebra ${\mathcal D}(A)\,:=\,{\mathcal D}\big(\,\odot(V^*)\,\big)$ of differential operators acting on the commutative algebra of polynomial functions on the affine space $A$ modeled on the vector space $V$ of finite dimension $n$ is the almost-commutative algebra of \textbf{polynomial differential operators} on the affine space $A$.
The Grothendieck algebra ${\mathcal D}(A)$ of polynomial differential operators is called the \textbf{Weyl algebra}. The Weyl algebra is  simple (\textit{i.e.} it has no nontrivial ideal) and central (\textit{i.e.} its center is the field $\mathbb K$ of the underlying commutative algebra $\mathcal A$), thus all its derivations are inner (\textit{i.e.} they arise from the adjoint action of some element of the algebra). 
The polynomial differential operators of order $k$ take the form $\hat{X}=\sum_{r=0}^k \frac1{r!}X^{a_1\cdots a_r}(y)\frac{\partial}{\partial y^{a_1}}\cdots\frac{\partial}{\partial y^{a_r}}$ with coefficients $X^{a_1\cdots a_r}(y)$ which are polynomials in $y$'s. The associated graded algebra $\text{gr}\, {\mathcal D}\big(\,{\mathbb R}[y^a]\,\big)$ is isomorphic to the commutative algebra ${\mathbb R}[y^a,p_b]$ of polynomials on the cotangent bundle $T^*{\mathbb R}^n\cong {\mathbb R}^n\oplus{\mathbb R}^{n*}$. 
}

\vspace{3mm}
\noindent{\small\textbf{Example (Formal differential operators)\,:} The Grothendieck algebra ${\mathcal D}\big(\,\overline{\odot}(V^*)\,\big)$ of differential operators acting on the commutative algebra $\overline{\odot}(V^*)$ of formal power series at the origin of the vector space $V$ of finite dimension $n$ is the almost-commutative algebra of \textbf{formal differential operators} at the origin of the vector space $V$.
A formal differential operator of order $k$ takes the form $\hat{X}=\sum_{r=0}^k \frac1{r!}X^{a_1\cdots a_r}(y)\frac{\partial}{\partial y^{a_1}}\cdots\frac{\partial}{\partial y^{a_r}}$ with coefficients $X^{a_1\cdots a_r}(y)=\sum_{q=0}^\infty \frac1{q!}X^{a_1\cdots a_r}_{b_1\cdots b_q}\,y^{b_1}\cdots y^{b_q}$ which are formal power series at the origin. The graded algebra $\text{gr} {\mathcal D}\big(\,{\mathbb R}\llbracket y^a\rrbracket \,\big)$ in the case of formal power series is isomorphic to the commutative algebra ${\mathbb R}\llbracket y^a,p_b]$ of formal power series in the Cartesian coordinates $y^a$ on the base but of polynomials in the vertical coordinates $p_b$ on $T^*{\mathbb R}^n$. In coordinate-independent terms, one has the isomorphism $\text{gr}{\mathcal D}({\mathcal A})\cong\overline{\odot}(V^*)\otimes \odot(V)$ of commutative algebras.
}

\vspace{3mm}
The Grothendieck algebra of differential operators acting on the commutative algebra $C^\infty(M)$ of smooth functions on a manifold $M$
is the almost-commutative algebra ${\mathcal D}(M)$ of smooth differential operators on the manifold $M$ endowed with the composition product $\circ$. Among other thing, it is an $\mathbb N$-filtered associative algebra, \textit{i.e.} the two following properties hold:
\be\label{Dinclusions}
{\mathcal C}^\infty(M)\cong{\mathcal D}^0(M)\subset {\mathcal D}^1(M)\subset {\mathcal D}^2(M) \subset \ldots \subset {\mathcal D}^k(M)\subset {\mathcal D}^{k+1}(M)\subset\ldots
\ee
and 
\be
{\mathcal D}^k(M)\circ {\mathcal D}^l(M)\subset {\mathcal D}^{k+l}(M)\,.
\ee
The above infinite sequence \eqref{Dinclusions} of inclusions provides an abstract definition of ${\mathcal D}(M)$ as the   direct limit.
Remember that the map 
\be
\hat{\bullet}\,:\,{\mathcal C}^\infty(M)\stackrel{\sim}{\to}{\mathcal D}^0(M)\,:\,f\mapsto\hat{f}\,,
\ee
which associates to any function $f$ its corresponding differential operator $\hat{f}:g\mapsto f\cdot g$ of order zero, is an isomorphism of commutative algebras mapping the unit function $1$ to the identity operator $\hat{1}$ and the pointwise product $f\cdot g$ of functions to the composition product $\hat{f}\circ\hat{g}$ of differential operators of order zero (\textit{i.e.} $\widehat{f\cdot g}=\hat{f}\circ\hat{g}$).

The commutator algebra $\mathfrak{D}(M)$ is the space ${\mathcal D}(M)$ of differential operators endowed with a structure of Lie algebra
through the commutator as Lie bracket $[\hat{X},\hat{Y}]$ between two differential operators $\hat{X}$ and $\hat{Y}$.\footnote{Sometimes, one will not specify which product structure (associative or Lie) is chosen and one will colloquially refer to it as the algebra of differential operators.}
Notice that 
\be\label{Dalmostcom}
[\,{\mathfrak{D}}^k(M)\,,\,{\mathfrak{D}}^l(M)\,]\subset {\mathfrak{D}}^{k+l-1}(M)\,.
\ee
As one can see, the first-order differential operators
span the non-abelian Lie subalgebra ${\mathfrak{D}}^1(M)$ while the
zeroth-order differential operators (\textit{i.e.} the functions) span the abelian Lie subalgebra ${\mathfrak{D}}^0(M)\cong  C^\infty(M)$.
The $\mathbb N$-filtered associative algebra ${\mathcal D}(M)$ is almost-commutative due to \eqref{Dalmostcom} and one has the following canonical isomorphisms
\be\label{isopol}
\text{gr}\,{\mathcal D}(M)\cong {\mathcal S}(M)\cong \Gamma(\odot TM)\,,
\ee
of Schouten algebras. In more concrete terms, there is a one-to-one correspondence between principal symbols 
\eqref{principsymb}
of differential operators and symmetric contravariant tensor fields $X^{\mu_1\ldots \mu_k}(x)$ relating the two commutative products (the pointwise product of functions on the cotangent space to the symmetrised product of symmetric contravariant tensor fields) and the two Poisson brackets (the canonical Poisson bracket of functions on the cotangent space to the Schouten bracket of symmetric contravariant tensor fields).

\subsubsection{Differential operators as almost-linear operators}\label{almostlinearops}

A commutative algebra ${\mathcal A}$ can be seen as a (bi)module over itself (via multiplication).
An ${\mathcal A}$-linear map from ${\mathcal A}$ (\textit{i.e.} an operator on ${\mathcal A}$ which is a morphism of ${\mathcal A}$-modules) to an ${\mathcal A}$-module is entirely determined by its action on the unit element $1\in {\mathcal A}$. In particular, the operator $\hat{f}=f\cdot$\, is the unique ${\mathcal A}$-linear map from ${\mathcal A}$ to itself, which maps the unit element to the function $f$, \textit{i.e.} such that $\hat{f}:1\mapsto f$. 

\vspace{5mm}\begin{figure}[h!]
\begin{framed}\noindent
\begin{center}
\textbf{Equivalent formulations of zeroth-order differential operators}
\end{center}

\noindent
Consider a commutative algebra ${\mathcal A}$. The following notions are equivalent: 
\begin{enumerate}

\item a differential operator $\hat{X}\in{\cal D}^0({\mathcal A})$ of order zero on ${\mathcal A}$, 

\item an endomorphism $\hat{X}\in\text{End}_{\mathcal A}({\mathcal A})$ of the ${\mathcal A}$-module ${\mathcal A}$, \textit{i.e.} an operator which is ${\mathcal A}$-linear in the sense that $\hat{X}[f\cdot g]=f\cdot\hat{X}[g]$ for any $f,g\in{\mathcal A}$, 

\item an endomorphism $\hat{X}\in\text{End}({\mathcal A})$ of the vector space ${\mathcal A}$ which is such that $\hat{X}\circ\hat{f}=\hat{f}\circ\hat{X}$ for any $f\in{\mathcal A}$.
\end{enumerate}
\vspace{3mm}
\end{framed}
\end{figure}
\vspace{5mm}

The definition of zeroth-order differential operators as $\mathcal A$-linear endomorphisms motivates a recursive definition of higher-order differential operators. 
A differential operator $\hat{X}\in{\mathcal D}^k({\mathcal A})$ on ${\mathcal A}$ of order $k$ can be equivalently defined as an endomorphism $\hat{X}\in\text{End}({\mathcal A})$ of the vector space ${\mathcal A}$ which will be said \textbf{almost} ${\mathcal A}$-\textbf{linear}, in the sense that $[\hat{X},\hat{f}]\in {\mathcal D}^{k-1}({\mathcal A})$ for any $f\in{\mathcal A}$, \textit{i.e.} it is ${\mathcal A}$-linear up to lower-order terms.
In other words, the Grothendieck algebra of differential operators acting on the commutative algebra ${\mathcal A}$ can be defined as the algebra of almost ${\mathcal A}$-linear operators on ${\mathcal A}$.

A ring ${\mathcal B}$ over a commutative algebra ${\mathcal A}$ such that the image of the unit map $i:{\mathcal A}\hookrightarrow{\mathcal B}$ lies in the center of $\cal B$, is called an $\mathcal A$-\textbf{algebra}. Any commutative ${\mathcal A}$-ring is an ${\mathcal A}$-algebra.

\vspace{3mm}
\noindent{\small\textbf{Example (Algebra over a field)\,:} In the particular case when ${\mathcal A}=\mathbb{K}$ is a field, a $\mathbb{K}$-algebra coincides with the usual notion of an associative algebra over $\mathbb K$.}
\vspace{3mm}

An $\mathbb N$-filtered algebra $\mathcal A$, which is such that its associated $\mathbb N$-graded algebra $\text{gr}{\mathcal A}$ is an ${\mathcal A}_0$-algebra over its component ${\mathcal A}_0$ of degree zero, will be called an \textbf{almost} $\mathcal{A}_0$-\textbf{algebra}. In terms of the commutator, the condition reads $[{\mathcal A}_i,{\mathcal A}_0]\subseteq {\mathcal A}_{i-1}$. Any almost-commutative algebra ${\mathcal A}$ is an almost ${\mathcal A}_0$-algebra, but the converse is not true in general.
The recursive definition of the Grothendieck algebra can be summarised as follows: the Grothendieck algebra ${\mathcal D}({\mathcal A})$ of differential operators on the commutative algebra $\mathcal A$ is the maximal subalgebra of the $\mathcal A$-ring $\text{End}({\mathcal A})$ of  endomorphisms of the vector space $\mathcal A$ which is an almost ${\mathcal A}$-algebra.

\subsubsection{Differential operators as infinitesimal automorphisms}

Let us now introduce the notation 
\be\label{defLieHS}
{\mathcal L}_{\hat X}f\,:=\,\hat{X}[f]\,.
\ee
Motivated by higher-spin gravity, it will be called the \textbf{higher-spin Lie derivative} ${\mathcal L}_{\hat X}f$ \textbf{of a function} $f$ \textbf{along a differential operator} $\hat{X}$ for reasons that will become clear later. 
The corresponding Lie algebra morphism (\textit{i.e.} 
$[{\mathcal L}_{\hat{X}},{\mathcal L}_{\hat{Y}}]={\mathcal L}_{[\hat{X},\hat{Y}]}$)
\be
{\mathcal L}\,:\,\mathfrak{D}(M)\,\to \mathfrak{gl}\Big(\, C^\infty(M)\,\Big)\,:\,{\hat X}\mapsto {\mathcal L}_{\hat X}
\ee
will be called the \textbf{fundamental representation of the algebra of differential operators}. The commutative algebra of functions is the representation space of this fundamental representation.
Let us stress the obvious point that the higher-spin Lie derivative along a differential operator is \textit{not} a derivation of the commutative algebra of functions, except when
the differential operator is a vector field (a tautology from the algebraic definition of vector fields as derivations), \textit{i.e.} ${\mathcal L}_{\hat X}\notin\mathfrak{der}\Big(\, C^\infty(M)\,\Big)$ for ${\hat X}\notin{\mathfrak{X}}(M)$. This is clear from the definition of the higher-spin Lie derivative but our choice of notation and terminology might obscure this point. Nevertheless, one will stick to this choice because it suggests a natural generalisation for the Lie derivative of differential operators. 

We define the \textbf{higher-spin Lie derivative} ${\mathcal L}_{\hat X} \hat{Y}$ \textbf{of a differential operator} $\hat{Y}$ \textbf{along another differential operator} $\hat{X}$ from compatibility with the Leibniz rule ${\mathcal L}_{\hat X}\big(\,\hat{Y}[f]\,\big)=({\mathcal L}_{\hat X} \hat{Y})[f]+\hat{Y}[{\mathcal L}_{\hat X}f]$, \textit{i.e.} we set
\ba
({\mathcal L}_{\hat X} \hat{Y})[f]&:=&{\mathcal L}_{\hat X}\big(\,\hat{Y}[f]\,\big)-\hat{Y}[{\mathcal L}_{\hat X}f]\\
&&\quad={\hat X}\big[\,\hat{Y}[f]\,\big]\,-\,{\hat Y}\big[\,\hat{X}[f]\,\big]\label{2ndlineHSLie}
\\
&&\quad=[\hat{X}\stackrel{\circ}{,}\hat{Y}]\,[f]\label{3rdlineHSLie}
\ea
where \eqref{defLieHS} was used to get \eqref{2ndlineHSLie}.
The last line \eqref{3rdlineHSLie}
results into the mere
identification of the higher-spin Lie derivative of a differential operator $\hat{Y}$ along the differential operator $\hat{X}$ with their commutator, \textit{i.e.} ${\mathcal L}_{\hat X} \hat{Y}=[\hat{X}\stackrel{\circ}{,}\hat{Y}]$. 
This generalisation of the Lie derivative leads to the following notation for the \textbf{adjoint representation of the algebra of differential operators} (on itself)
\be
{\mathcal L}\,:\,\mathfrak{D}(M)\,\to \mathfrak{inn}\Big(\,{\mathcal D}(M)\,\Big)\,:\,{\hat X}\mapsto {\mathcal L}_{\hat X}
\ee
via inner derivations of the almost-commutative algebra of differential operators. Locally, all derivations of the almost-commutative algebra ${\mathcal D}(M)$ of differential operators are inner derivations.\footnote{To be more precise, the number of linearly independent outer derivations of the almost-commutative algebra of differential operators on the manifold $M$ is equal to the first Betti number of $M$ \cite{Grabowski:2003b}.} Therefore, these higher-spin Lie derivatives essentially exhaust all infinitesimal symmetries of the almost-commutative algebra of differential operators. The associative algebra of differential operators is comparable to the Lie algebra of vector fields: they are both self-referential objects in that they coincide with their own collection of infinitesimal symmetries. Strictly speaking, this is only true locally for ${\mathcal D}(M)$ while this is true globally for ${\mathfrak{X}}(M)$.

Let us stress an important subtlety and potential source of confusion related to the fact that any function $f\in C^\infty(M)$ can be seen as a differential operator $\hat{f}\in{\mathcal D}^0(M)$ of order zero. As a differential operator, $\hat{f}$ transforms into the adjoint representation, its higher-spin Lie derivative along ${\hat X}\in{\mathcal D}^k(M)$ is ${\mathcal L}_{\hat X}\hat{f}=[\hat{X},\hat{f}]\in{\mathcal D}^{k-1}(M)$. When ${\hat X}$ is a higher-order differential operator, then the higher-spin Lie derivative of $\hat{f}$ along $\hat{X}$ is \textit{not} a differential operator of order zero, \textit{i.e.} ${\mathcal L}_{\hat X}\hat{f}\notin{\mathcal D}^{0}(M)$ in general. However, one may extract a function by acting on the unit element. More precisely, the relation between the fundamental and adjoint representation is as follows: ${\mathcal L}_{\hat X}f=({\mathcal L}_{\hat X}\hat{f})[1]$. 

\subsubsection{No-go theorem against (naive) higher-spin diffeomorphisms}\label{formalexpdiffop}

The almost-commutative algebra of differential operators and the Schouten algebra of principal symbols admit somewhat few finite symmetries, while there is a plethora of infinitesimal symmetries (see Section 7 of \cite{Grabowski:2002} and Section 8 of \cite{Grabowski:2003b} for more details).
The finite automorphisms of the almost-commutative algebra ${\mathcal D}(M)$ of differential operators are very scarce with respect to the infinitesimal automorphisms. In fact, the finite automorphisms of the almost-commutative algebra ${\mathcal D}(M)$ essentially coincide with internal Abelian gauge symmetries (as in Maxwell electromagnetism) and standard diffeomorphisms of the manifold $M$ (as in general relativity), for which the commutative subalgebra ${\mathcal D}^0(M)$ of functions is an invariant subspace. While the infinitesimal automorphisms (the higher-spin Lie derivative) are perfectly well defined (and just a fancy name for the adjoint action of differential operators) however their formal exponentiation is not (see \textit{e.g.} \cite[Sect.2]{Bekaert:2021sfc} for more details) in general, except for the inner automorphisms generated by first-order differential operators.

This prevents a too naive definition of ``higher-spin diffeomorphisms'' (attempting to overcome this obstacle was the focus of \cite{Bekaert:2021sfc}). One may expect these subtleties to be related to the elusive (non)locality properties of higher-spin gravity.
Let us stress that this feature is not specific to the almost-commutative algebra ${\mathcal D}(M)$ of differential operators, the same holds for the automorphisms of the Schouten algebra ${\mathcal S}(M)$ of symbols. 

\vspace{3mm}
\noindent{\textbf{Theorem (Grabowski \& Poncin)\,:}
\textit{Any one-parameter group of automorphisms of the associative algebra ${\mathcal D}(M)$ of differential operators (respectively, of the symplectic algebra ${\mathcal S}(M)$ of symbols) is generated by a first-order differential operator (respectively, by a symbol of degree one).}}
\vspace{3mm}

By contraposition, this result of \cite{Grabowski:2002} can be expressed equivalently as a no-go theorem.

\vspace{3mm}
\noindent\textbf{No-go theorem\,:} \textit{Higher-spin Lie derivatives along a higher-order differential operator on $M$ (respectively, higher-degree Hamiltonian vector fields on $T^*M$) cannot be integrated to one-parameter groups of automorphisms of the associative algebra ${\mathcal D}(M)$ of differential operators (respectively, of the symplectic algebra ${\mathcal S}(M)$ of symbols).}
\vspace{3mm}

Concretely, the obstruction is that the polynomiality in the momenta is \textit{not} preserved by the formal exponentiation of symplectic vector fields on $T^*M$, even if this symplectic vector field is itself polynomial in the momenta. This problem was addressed at length in \cite{Bekaert:2021sfc} so it will not be reviewed here. 
In any case, not much is actually known about automorphism groups of algebras of differential operators. These infinite-dimensional groups have surprising properties and remain difficult mathematical objects of current study, even in the polynomial case, as examplified by the following two conjectures.

\vspace{3mm}
\noindent{\small\textbf{Example (Kontsevitch conjecture)\,:} It was argued in 2005 that the automorphism group of the Weyl algebra ${\mathcal D}(A)$, the associative algebra of polynomial differential operators on a finite-dimensional affine space $A$, must be isomorphic to the automorphism group of the associated Poisson algebra ${\mathcal S}(A)$ of polynomial symbols \cite{KanelBelov:2005}, 
\be\label{isomDASA}
Aut\big(\,{\mathcal D}(A)\,\big)\,\cong\,Aut\big(\,{\mathcal S}(A)\,\big)\,,
\ee 
but it was only in 2018 that a complete proof of this conjecture was outlined \cite{KanelBelov:2018}. Note that, by construction, the automorphism group of the Poisson algebra of polynomial symbols is isomorphic to the group of polynomial symplectomorphisms (\textit{i.e.} polynomial diffeomorphisms preserving the symplectic structure of the cotangent bundle $T^*A$). 
The isomorphism \eqref{isomDASA} may come as a surprise because it is not true at the level of infinitesimal automorphisms: the Lie algebra of derivations of the associative algebra ${\mathcal D}(A)$ of polynomial differential operators is \textit{not} isomorphic to the Lie algebra of derivations of the Poisson algebra ${\mathcal S}(A)$ of polynomial symbols,
\be
\mathfrak{der}\big(\,{\mathcal D}(A)\,\big)\,\not\cong\,\mathfrak{der}\big(\,{\mathcal S}(A)\,\big)\,
\ee 
although they are isomorphic as vector spaces \cite{KanelBelov:2005}.\footnote{Note that all derivations of the Poisson algebra ${\mathcal S}(A)$ and of the associative algebra ${\mathcal D}(A)$ are inner (hence they are polynomial Hamiltonian vector fields and higher-spin Lie derivatives along polynomial differential operators, respectively).
}}

\vspace{3mm}
\noindent{\small\textbf{Example (Dixmier conjecture)\,:} In 1968, Dixmier asked \cite{Dixmier:1968} if any endomorphism of the Weyl algebra of polynomial differential operators on a finite-dimensional affine space $A$ is an automorphism. In other words, whether any endomorphism of the Weyl algebra is invertible. A positive answer to this question is referred to as the Dixmier conjecture. 
This conjecture is equivalent to its Poisson counterpart (sometimes called the Poisson conjecture \cite{Adjamagbo})\,: that any endomorphism of the Poisson algebra of polynomial symbols is an automorphism.
If the Dixmier conjecture is true, this means that the endomorphism algebra $\text{End}\big(\,{\mathcal D}(A)\,\big)$ and the automorphisms group $Aut\big(\,{\mathcal D}(A)\,\big)$ of the Weyl algebra are isomorphic, as \textit{associative algebras}.
}

\vspace{3mm}
Let us summarise again the situation in the smooth case: 
Firstly, if one requires to preserve the filtration/graduation then the only inner derivations of the almost-commutative algebra ${\mathcal D}(M)$ of differential operators (respectively, of the Schouten algebra ${\mathcal S}(M)$ of principal symbols) which are integrable to one-parameter groups of automorphisms of the same filtered (respectively, graded) algebra(s), are the (complete) Lie derivatives along vector fields on $M$. Secondly, even if one does not require \textit{a priori} to preserve the filtration/graduation, the only inner derivations of the associative algebra ${\mathcal D}(M)$ of differential operators (respectively, of the Poisson algebra ${\mathcal S}(M)$ of symbols) which are integrable to one-parameter groups of automorphisms of the same associative (respectively, Poisson) algebra(s), are again the (complete) Lie derivatives along vector fields on $M$. In other words, the automorphisms of these associative/Poisson algebras must necessarily preserve their filtration/graduation and, consequently,
higher-spin Lie derivatives along higher-derivative differential operators (respectively, Hamiltonian vector fields with higher-degree symbols as Hamiltonians) are \textit{not} integrable to one-parameter groups of automorphisms of the associative algebra ${\mathcal D}(M)$ of differential operators (respectively, of the Poisson algebra ${\mathcal S}(M)$ of symbols).\footnote{A way out is that these inner derivations \textit{can} be integrated to one-parameter groups of automorphisms of larger algebras, see \textit{e.g.} the two proposals in \cite{Bekaert:2021sfc}.} If a Lie algebra\footnote{Note that the so-called ``Lie's third theorem'' stating that every finite-dimensional real Lie algebra $\mathfrak g$ is associated to a real Lie group $G$ does not hold in general for infinite-dimensional Lie algebras. } is integrable to a Lie group, then one would expect (for any reasonable topology) that all its inner derivations are integrable to one-parameter groups of automorphisms of the Lie algebra (via the exponential map).
Accordingly, an important corollary of these negative results is the strong no-go theorem of Grabowski and Poncin \cite[Cor.4]{Grabowski:2002}:

\vspace{3mm}
\noindent\textbf{No-go theorem (Grabowski \& Poncin)\,:} \textit{The two infinite-dimensional Lie algebras, ${\mathfrak{D}}(M)$ of differential operators and ${\mathcal S}(M)$ of symbols on a manifold $M$, are not integrable to (infinite-dimensional) Lie groups (of which they are the Lie algebras).}
\vspace{3mm}

This no-go theorem may be responsible for the elusive locality properties of higher-spin interactions. In any case, it makes manifest that if finite higher-spin symmetries are taken seriously, then they require to leave the realm of operators with bounded number in the derivatives (the landmark of locality). A toy-model example \cite[Sect.6]{Bekaert:2021sfc} of completion of the almost-commutative algebra of differential operators bypassing the no-go theorem of Grabowski and Poncin is reviewed in the next subsection.

\subsubsection{Yes-go theorem on (formal) higher-spin diffeomorphisms}\label{almostdiff}

One simple trick to go beyond differential operators is to make use of a formal deformation parameter, say $\hbar$.

Let $V$ be a vector space over the field $\mathbb K$. Then $V\llbracket\hbar\rrbracket$ denotes the vector space of formal power series in $\hbar$ with coefficients that are elements of $V$. 
The vector space $V\llbracket\hbar\rrbracket$ is a ${\mathbb{K}}\llbracket\hbar\rrbracket$-module. A $\mathbb{K}\llbracket\hbar\rrbracket$-linear map $U\in\text{End}_{\mathbb{K}\llbracket\hbar\rrbracket}\Big(\,V\llbracket\hbar\rrbracket\,\Big)$ is said \textbf{$\hbar\,$-linear} and is uniquely determined from its restriction $T=U|_V\,:\,V\to V\llbracket\hbar\rrbracket$ to the subspace $V\subset V\llbracket\hbar\rrbracket$ of power series independent of $\hbar$. This restriction is a ${\mathbb{K}}$-linear map from $V$ to $V\llbracket\hbar\rrbracket$, hence it can be thought of as an element of the space $\text{End}(V)\llbracket\hbar\rrbracket$ of formal power series in $\hbar$ with coefficients that are endomorphisms of the vector space $V$. This leads to the isomorphism $\text{End}_{\mathbb{K}\llbracket\hbar\rrbracket}\Big(\,V\llbracket\hbar\rrbracket\,\Big)\cong\text{End}(V)\llbracket\hbar\rrbracket$ of associative algebras. With a slight abuse of notation, the ${\mathbb{K}}$-linear map $T$ and its unique $\hbar$-linear extension $U$ will be denoted by the same symbol from now on.

Let us assume that $V$ is $\mathbb N$-filtered and denote by $V\,\llangle\hbar\rrangle$ the vector space spanned by formal power series in $\hbar$ 
with coefficients which are of degree \textit{smaller or equal} to the power of $\hbar$, 
\be
v(\hbar)\in V\,\llangle\hbar\rrangle\quad\Longleftrightarrow\quad v(\hbar)\,=\,\sum\limits_{n=0}^\infty v_n\,\hbar^n\quad\text{with}\quad v_n\in V_n\,.
\ee
It will be called the \textbf{$\hbar$-filtered extension of the $\mathbb N$-filtered space} $V$. 
Let us now assume that $V$ is $\mathbb N$-graded and denote by $V\,\lVert\hbar\rVert$ the vector space spanned by formal power series in $\hbar$ with coefficients which are of grading \textit{equal} to the power of $\hbar$.
It will be called the \textbf{$\hbar$-graded extension of the $\mathbb N$-graded space} $V$.

Consider an associative algebra $\mathcal A$\,. The $\hbar$\,-linear extension of its product endows the vector space $\mathcal{A}\llbracket\hbar\rrbracket$ with a structure of $\mathbb{K}\llbracket\hbar\rrbracket$-algebra. 
Furthermore, if the associative algebra is $\mathbb N$-filtered (respectively, $\mathbb N$-graded) then the $\hbar$\,-linear extension of the product of the associative algebra $\mathcal{A}$ endows the vector space $\mathcal{A}\,\llangle\hbar\rrangle\subset\mathcal{A}\llbracket\hbar\rrbracket$ (respectively, $\mathcal{A}\,\lVert\hbar\rVert\subset\mathcal{A}\llbracket\hbar\rrbracket$) with a structure of associative (sub)algebra.

The elements of the form $a(\hbar)=\hbar\,\,b(\hbar)$, where $b(\hbar)\in\mathcal{A}\llbracket\hbar\rrbracket$\,, form a proper ideal $\hbar\,\mathcal{A}\llbracket\hbar\rrbracket\subset\mathcal{A}\llbracket\hbar\rrbracket\,$.
Similarly, the elements of the same form but where $b(\hbar)\in\mathcal{A}\,\llangle\hbar\rrangle$\,, form a proper ideal of $\hbar\,\mathcal{A}\,\llangle\hbar\rrangle\subset\mathcal{A}\,\llangle\hbar\rrangle$\,.
The quotient algebra 
of the $\hbar$-filtered extension $\mathcal{A}\,\llangle\hbar\rrangle$ of the 
filtered associative algebra $\mathcal A$ by the ideal $\hbar\,\mathcal{A}\,\llangle\hbar\rrangle$
is isomorphic to the
$\hbar$-graded extension of the 
associated $\mathbb N$-graded algebra $\mathcal{B}=\text{gr}\mathcal{A}$,
\be\label{quotalg}
\mathcal{A}\,\llangle\hbar\rrangle\,/\,\,\hbar\,\mathcal{A}\,\llangle\hbar\rrangle\,\cong\,\text{gr}\mathcal{A}\,\lVert\hbar\rVert\,.
\ee

An $\mathbb N$-filtered associative algebra $\mathcal{A}$ is almost-commutative iff its $\hbar$-filtered extension $\mathcal{A}\,\llangle\hbar\rrangle$ is commutative modulo $\hbar$, 
\be
\big[\,\mathcal{A}\,\llangle\hbar\rrangle\,,\,\mathcal{A}\,\llangle\hbar\rrangle\,\big]\subseteq \hbar\,\mathcal{A}\,\llangle\hbar\rrangle\,.
\ee
which is equivalent to say that the quotient algebra \eqref{quotalg} is commutative.
A Poisson algebra $\mathcal{B}$ is a Schouten algebra iff it is an $\mathbb N$-graded vector space such that $\mathcal{B}\,\lVert\hbar\rVert$ is an algebra for the $\hbar$-linear extension of the product of the commutative algebra $\mathcal{B}$ and the $\hbar$-linear extension of the Poisson bracket obeys
\be
\big\{\,\mathcal{B}\,\lVert\hbar\rVert\,,\,\mathcal{B}\,\llangle\hbar\rrangle\,\big\}\subseteq\hbar\,\mathcal{B}\,\lVert\hbar\rVert\,.
\ee
The quotient algebra of the $\hbar$-filtered extension $\mathcal{A}\,\llangle\hbar\rrangle$ of an almost commutative algebra $\mathcal A$ by the ideal $\hbar\,\mathcal{A}\,\llangle\hbar\rrangle$
is endowed with a structures of Poisson algebra via the commutative product induced from the associative product and via the Poisson bracket induced from $\frac1{\hbar}[\,\,,\,]$. Furthermore, this quotient is isomorphic to the $\hbar$-graded extension of the associated Schouten algebra $\mathcal{B}=\text{gr}\mathcal{A}$.

Let $\mathcal A$ be a commutative algebra. The $\mathbb{K}\llbracket\hbar\rrbracket$-subalgebra 
$\mathcal{B}\subset\text{End}(\mathcal{A})\llbracket\hbar\rrbracket$ of $\hbar$-linear endomorphisms of the vector space $\mathcal{A}\llbracket\hbar\rrbracket$ which are $\mathcal{A}$-linear modulo $\hbar$, \textit{i.e.}
\be\label{hbarmodulolinear}
\big[\,\mathcal{B}\,\llangle\hbar\rrangle\,,\,\mathcal{D}^0(\mathcal{A})\,\big]\subset\hbar\,\mathcal{B}\,\llangle\hbar\rrangle\,\,,
\ee
is the $\hbar$-filtered extension $\mathcal{D}(\mathcal{A})\llangle\hbar\rrangle$ of the Grothendieck algebra of differential operators on $\mathcal A$, as can be checked by spelling out the condition \eqref{hbarmodulolinear} in powers of $\hbar$.

Remember that for any associative algebra $\mathcal{A}$, one can form the Lie algebra $\mathfrak{A}$ by endowing the vector space $\mathcal{A}$ with the commutator as Lie bracket.
Consider an almost-commutative algebra $\mathcal{A}$, one can define the following representation of the Lie algebra $\mathfrak{A}\,\llangle\hbar\rrangle$ on the associative algebra $\mathcal{A}\,\llangle\hbar\rrangle$
\be
ad^\hbar\,:\,\mathfrak{A}\,\llangle\hbar\rrangle\to\mathfrak{der}\big(\,\mathcal{A}\,\llangle\hbar\rrangle\,\big)\,:\,b\mapsto ad^\hbar_b
\ee
where 
\be
ad^\hbar_b\,:=\,\frac1{\hbar}\,ad_b
\ee
is an $\hbar\,$-linear derivation of the associative algebra $\mathcal{A}\,\llangle\hbar\rrangle$\,. 

\vspace{3mm}
\noindent{\small\textbf{Technical Lemma \cite{Bekaert:2021sfc}\,:} Consider an almost-commutative algebra $\mathcal A$. Let us assume that the adjoint action $ad|_{\mathfrak{A}_1}\,:\,\mathfrak{A}_1\to\mathfrak{der}(\mathcal{A})$ of the Lie subalgebra $\mathfrak{A}_1\subset\mathfrak{A}$ of degree one is integrable in the sense that there are one-parameter groups of automorphisms of the almost-commutative algebra $\mathcal A$ with elements of the form $\exp(\,t\, ad_a)\in Aut(\mathcal{A})$ for some $a\in\mathcal{A}_1$. Then the action $ad^\hbar$ of the Lie algebra $\mathfrak{A}\,\llangle\hbar\rrangle$ on the associative algebra $\mathcal{A}\,\llangle\hbar\rrangle$ is essentially integrable, in the sense that all elements $a(\hbar)=\sum_{n=1}^\infty a_n\,\hbar^n\in \mathcal{A}\,\llangle\hbar\rrangle\cap\hbar\,\mathcal{A}\llbracket\hbar\rrbracket$ generate one-parameter groups of automorphisms of the associative algebra $\mathcal A$ of the form $\exp(\,t\, ad^\hbar_a)\in Aut(\mathcal{A}\,\llangle\hbar\rrangle)$ when the coefficient $a_1$ is integrable in the previous sense.
}
\vspace{3mm}

This lemma applies in particular for the Grothendieck algebra of differential oeprators on any commutative algebra. Therefore, a direct corollary is that the no-go theorem of Grabowski and Poncin can be bypassed by considering instead the $\hbar$-filtered completion $\mathcal{D}(M)\,\llangle\hbar\rrangle$ of the associative algebra $\mathcal{D}(M)$ of differential operators.
It is spanned by formal power series in $\hbar$ with coefficients that are differential operators on $M$ of order smaller or equal to the power of $\hbar$,
\be\label{Xhbar}
\hat{X}_\hbar\,=\,\sum\limits_{r=0}^\infty\hat{X}_r\,\hbar^r\,,\qquad\hat{X}_r\in\mathcal{D}^r(M)\,.
\ee
They were called \textbf{almost-differential operators} on the manifold $M$ in \cite{Bekaert:2021sfc}.  

\vspace{3mm}
\noindent{\textbf{Yes-go proposition\,:} \textit{Any almost-differential operator $\hat{X}_\hbar\in\mathcal{D}(M)\,\llangle\hbar\rrangle\cap\hbar\,\mathcal{D}(M)\llbracket\hbar\rrbracket$ 
is locally integrable to a one-parameter group of automorphisms of the algebra $\mathcal{D}(M)\llangle\hbar\rrangle$ of almost-differential operators.}
}

\subsection{Higher-order jets}

\subsubsection{Higher-order contact ideals}

Consider a point $m$ of $M$ and let $k$ be a non-negative integer (or infinity). 
The commutative ideal ${\mathcal I}^k(m)$ of $ C^\infty(M)$
is spanned by the functions $f$ such that $(\partial_{\mu_1}\ldots\partial_{\mu_r} f)|_m=0$ for $0\leqslant r\leqslant k$ (respectively, for all $r$ when $k=\infty$).
It will be called the \textbf{contact ideal of order} $k$ \textbf{at the point} $m$.
Notice that, although one would make use of a specific coordinate system to write down these partial derivatives, the commutative ideal 
${\mathcal I}^k(m)$
does not depend on the choice of coordinates, as can be checked explicitly. Therefore, all the following notions will also be coordinate-free,
although this may not be obvious at first sight.
Since one deals with smooth functions, one may easily check that the order-$k$ contact ideal is identical to the $(k+1)$-fold pointwise power of the maximal ideal
${\mathcal I}^0(m)$:
\be
{\mathcal I}^k(m)\,=\,\Big(\,{\mathcal I}^0(m)\,\Big)^{k+1}\,.\label{powerideal}
\ee
While the maximal ideal
${\mathcal I}^0(m)$ captures algebraically the geometric notion of the point $m$ on the manifold, the commutative ideal ${\mathcal I}^k(m)$
captures something more refined: the infinitesimal vicinity of the point $m$ till order $k$. 

One can see the structure algebra as the contact ideal of order $-1$, \textit{i.e.} introduce the notation  ${\mathcal I}^{-1}(m):= C^\infty(M)$ in order to have the following infinite sequence of canonical inclusions
\be
\ldots \hookrightarrow {\mathcal I}^{k+1}(m)\hookrightarrow {\mathcal I}^k(m)
\hookrightarrow\ldots\hookrightarrow {\mathcal I}^1(m) \hookrightarrow {\mathcal I}^0(m)
\hookrightarrow {\mathcal I}^{-1}(m)\,,
\label{icontactid}
\ee
of vector spaces whose inverse limit defines the contat ideal ${\mathcal I}^\infty(m)$ of infinite order at $m$. Therefore $ C^\infty(M)$ is a $\mathbb Z$-filtered vector space, with decreasing filtration, and the corresponding $\mathbb N$-graded vector space is isomorphic to the symmetric tensor product $\odot T^*_mM$ of the cotangent space, \textit{i.e.} 
\be\label{cotangentcontactideals}
\odot T^{*k+1}_mM\,\cong\,{\mathcal I}^k(m)\,/\,{\mathcal I}^{k+1}(m)\,.
\ee

The contact ideal ${\mathcal I}^k(m)$ of order $k$ is a commutative ideal of the structure algebra $ C^\infty(M)$ so it defines an equivalence relation among functions $f$, whose equivalence classes are denoted by $j^k_m f$ and are called \textbf{jets of order} $k$ \textbf{at the point} $m$
(or simply $k$-jets) of functions.
By definition, two functions $f$ and $g$ define the same $k$-jet $j^k_m f$ at $m$ 
iff all their derivatives till order $k$ have the same
value at $m$.

The quotient 
\be\label{JkM}
J^k_mM\,:=\, C^\infty(M)\,/\,{\mathcal I}^k(m)
\ee
is the $k$-\textbf{jet space} at $m$.  The infinite-jet space $J^\infty_mM$ will often be called simply ``jet space at $m$'' for short. 
Since the $k$-jet spaces are defined as the quotient of a commutative algebra by an ideal, they are commutative algebras as well. In order to emphasise this property, they will sometimes be referred to as $k$-\textbf{jet space algebras}. However, they will be more often called jet spaces because it is their structure of vector spaces which is usually the most relevant.
 
\subsubsection{Truncated polynomials and formal power series}

Let $V$ be a finite-dimensional vector space. Consider the symmetric algebra $\odot(V^*)$ (thought as the commutative algebra of polynomials on $V$) with maximal idea $\mathfrak{m}=\odot^{>0}(V^*)$ of symmetric tensors of non-vanishing rank (thought as the contact ideal ${\mathcal I}(0)$ of polynomials vanishing at the origin). Its $(k+1)$-th power $\mathfrak{m}^{k+1}\cong\odot^{>k}V^*$ is spanned by symmetric tensors of rank higher than $k$ (thought as the $k$-contact ideal ${\mathcal I}^k(0)$ of polynomials vanishing till order $k$ at the origin).
The quotient 
\be
J^k_0V\,:=\,\odot (V^*)\,/\,\odot^{>k}(V^*)
\ee
is the $k$-jet space algebra at the origin.
The (inverse) limit $J^\infty_0V$ of these quotients for $k=\infty$ is a local algebra which can be thought of as the commutative algebra $\overline{\odot}(V^*)$ of formal power series at the origin of $V$. A theorem of Borel states that any formal power series is the Taylor series of a smooth function \cite{Borel}. Therefore, the commutative algebra of formal power series at the origin of the vector space $V$ is isomorphic to the commutative algebra of smooth functions on $V$ quotiented by the contact ideal of infinite order
\be
J^\infty_0V\,:=\,{\mathcal C}^\infty(V)\,/\,{\mathcal I}^\infty(0)\,.
\ee

A finite-dimensional local algebra $\mathcal A$ is called a \textbf{Weil algebra} \cite{Weil} (see also \cite[Sec.35]{Kolar}). Any Weil algebra $\mathcal A$ is a direct sum ${\mathcal A}=\mathbb{R}\oplus\cal N$, where the maximal ideal ${\mathcal N}\subset{\mathcal A}$ is nilpotent (\textit{i.e.} ${\mathcal N}^k=0$ for some finite $k\in\mathbb N$). For any Weil algebra $\mathcal A$, there exists a finite-dimensional vector space $V$ such that $\mathcal A$ is a quotient of the local algebra $\overline{\odot}(V^*)\cong J^\infty_0V$ of formal power series at the origin, by a commutative ideal ${\mathcal I}$ of finite codimension. 

\vspace{3mm}
\noindent{\small\textbf{Example (Truncated polynomials)\,:} The $k$-jet space algebra $J_0^kV$ at the origin of the vector space $V$ is the quotient $J^\infty_0V\,/\,{\mathcal I}^{k+1}(0)$ of the local algebra $J_0^\infty V$ of power series at the origin by the $(k+1)$-fold power of its maximal ideal ${\mathcal I}(0)$.  The $k$-jet space algebras $J_0^kV$ are important examples of Weil algebras. Its elements are equivalence classes of power series modulo terms of order $k+1$, which are sometimes called \textbf{truncated polynomials} of degree $k$.
In order to see this more concretely, let us introduce Cartesian coordinates $\varepsilon^\mu$ on ${\mathbb R}^n$, then the $k$-jet space algebra $J_0^k{\mathbb R}^n$ at the origin of ${\mathbb R}^n$ is isomorphic to the commutative algebra of truncated polynomials of degree $k$ in the variable $\varepsilon^\mu$, \textit{i.e.} power series modulo terms of order $k+1$,
\be
J_0^k{\mathbb R}^n\,\cong\,{\mathbb R}\llbracket \varepsilon^\mu\rrbracket \,/\,\Big(\,\mathfrak{m}\big(\,{\mathbb R}\llbracket \varepsilon^\mu\rrbracket \,\big)\,\Big)^{k+1}\,.
\ee
}

\vspace{1mm}
A convenient representation for a $k$-jet at $m$ of a smooth function is as the Taylor expansion of degree $k$ (for $k=\infty$, this representation is actually a formal power series). It is uniquely defined by all its derivatives at $m$ till order $k$.
A compact expression for a $k$-jet is as an equivalence class of power series $\phi(\varepsilon)$ in the auxiliary variable $\varepsilon^\mu$: 
\be\label{kjetpowerser}
\phi(\varepsilon)+{\mathcal O}\big(\,|\varepsilon|^{k+1}\,\big) \,=\,\sum\limits_{r=0}^k \frac1{r!}\,\,\phi_{\mu_1\ldots \mu_r}(x)\,\varepsilon^{\mu_1}\ldots \varepsilon^{\mu_r}\quad\text{modulo}\quad|\varepsilon|^{k+1} \,.
\ee
For $k=\infty$, the extra term (``modulo ...'') can be consistently dropped in the previous expression.
The equivalence class \eqref{kjetpowerser} can be seen as a Taylor series at $m$ modulo terms of order $\varepsilon^{k+1}$.
The product is well defined on the above equivalence classes (truncated polynomials) but is not preserved by a choice of representatives (genuine polynomials) since the product of two polynomials of degree $k$ is a polynomial of degree $2k$ rather than $k$. 

\subsubsection{Higher-order jet bundles}

The $k$-\textbf{jet bundle} is the manifold $J^kM=\bigcup_m J^k_m M $ and it has local coordinates 
$$(\,x^\nu\,,\,\phi\,,\,\phi_\mu\,,\,\phi_{\mu_1\mu_2}\,,\,\ldots\,,\,\phi_{\mu_1 \cdots\,\mu_k}\,)\,.$$
The transformation law of a $k$-jet under the coordinate transformation $x^\mu\mapsto x^{\prime\mu}(x)$ is as follows:
\be
\phi^\prime=\phi\,,\quad \phi_\mu^\prime=\frac{\partial x^\nu}{\partial x^{\prime\mu}}\,\phi_\nu\,,\quad 
\phi_{\mu_1\mu_2}^\prime=\frac{\partial x^{\nu_1}}{\partial x^{\prime\mu_1}}\,\frac{\partial x^{\nu_2}}{\partial x^{\prime\mu_2}}
\,\phi_{\nu_1\nu_2}+\frac{\partial^2 x^\nu}{\partial x^{\prime\mu_1}\partial x^{\prime\mu_2}}\,\phi_\nu\,,\quad\cdots
\label{coordtransfojet}
\ee
As one can see, the component $\phi_{\mu_1 \cdots\,\mu_r}$ with $r$ indices does not transform as a rank-$r$ symmetric covariant tensor because the lower rank components contribute in the transformation law of a given component of a $k$-jet.
In other words, only the collection of all its non-trivial components $\phi_\mu$, $\phi_{\mu_1\mu_2}$, ..., $\phi_{\mu_1 \cdots\,\mu_k}$ is
a coordinate-free object (the rank-zero can of course be separated).

There is an infinite tower of vector bundles fibrations:
\be
\ldots \twoheadrightarrow J^{k+1}M\twoheadrightarrow J^kM
\twoheadrightarrow\ldots\twoheadrightarrow J^2M \twoheadrightarrow J^1M
\twoheadrightarrow J^0M\cong M\times{\mathbb R}\,,
\label{pjetb}
\ee
the inverse limit of which provides an abstract definition of the infinite-jet bundle $J M$.\footnote{Although it is very tempting, one will refrain from calling the set of jets $J M$ the ``jet set'' of the manifold $M$.} The kernel of the projection $J^{k}M\twoheadrightarrow J^{k-1}M$ is the bundle of symmetric covariant tensors of rank $k$, i.e. $\odot^kT^*M$, in agreement with the previous isomorphism for the quotients of contact ideals.

The sections of the $k$th jet bundle $J^kM$ will be called 
$k$-\textbf{jet fields}. 
The commutative algebra of $k$-jet fields will be denoted $\mathcal{J}^k(M):=\Gamma(J^kM)$.
As mentioned above, a convenient representative for a $k$-jet is a truncated polynomial of degree $k$ (or a formal power series for $k=\infty$). A $k$-jet field $\phi$ is therefore compactly written as a function of two variables
\be\label{kjetfield}
\phi(x;\varepsilon)+{\mathcal O}(|\varepsilon|^{k+1}) \,=\,\sum\limits_{r=0}^k \frac1{r!}\,\,\phi_{\mu_1\ldots \mu_r}(x)\,\varepsilon^{\mu_1}\ldots \varepsilon^{\mu_r}\quad\text{modulo}\quad|\varepsilon|^{k+1} \,.
\ee
A very important point is that, for a generic $r$, the coefficient $\phi_{\mu_1\ldots \mu_r}(x)$ in the $k$-jet field $\phi(x;\varepsilon)$
is \textit{not} a symmetric covariant tensor fields of rank $r$, although the notation may suggest this misleading interpretation. 
This subtlety is more obvious if ones considers
the $k$th \textbf{prolongation} of a function $f$ that is the section $j^kf$ of the $k$th jet bundle $J^kM$ whose local expression is 
the Taylor expansion of order $k$:
\begin{eqnarray}
(j^kf)(x;\varepsilon) &=&\sum\limits_{r=0}^k \frac1{r!}\,\, \partial_{\mu_1}\ldots\partial_{\mu_r} f(x)\,\varepsilon^{\mu_1}\ldots \varepsilon^{\mu_r} \quad\text{modulo}\quad|\varepsilon|^{k+1} 
\label{Taylork}\\
&=&f(x+\varepsilon)\quad\text{modulo}\quad|\varepsilon|^{k+1} \,,
\end{eqnarray}
where the last equality holds for any finite $k$ by virtue of Taylor's theorem. Obviously, the coefficients $\partial_{\mu_1}\ldots\partial_{\mu_r} f(x)$ do \textit{not} transform as symmetric covariant tensor fields of rank $r$ under general coordinate transformations.
Notice that the $k$th prolongation is a linear map $j^k: C^\infty(M)\to \mathcal{J}^k(M)$ which has the following concrete representation in coordinates
\be
j^k\,=\,\exp\left(\varepsilon^\mu\frac{\partial}{\partial x^\mu}\right)\,+\,{\mathcal O}\big(\,|\varepsilon|^{k+1}\,\big)\,,
\ee
in agreement with \eqref{Taylork}.
The infinite prolongation $j^\infty: C^\infty(M)\to {\mathcal J}^\infty(M)$ plays an important role in the theory of partial differential equations (see \textit{e.g.} the textbook \cite{Krasilshchik}) and has a particularly simple representation $j^\infty=\exp(\varepsilon^\mu\frac{\partial}{\partial x^\mu})$ in local coordinates as a formal power series in $\varepsilon$.

\vspace{3mm}
\noindent{\small\textbf{Remark:} For analytic functions, their infinite prolongation can formally be understood as a mere translation: $(j^\infty f)(x;\varepsilon) =f(x+\varepsilon)$\,. However, this equality is not correct for generic smooth functions
since the Taylor series of a smooth function at a point does not necessarily converge, and even if it converges 
this Taylor series is not necessarily equal to the value of the function at this point.
Along the same lines, one should always keep in mind the theorem according to which there is an
infinite collection of smooth functions defining the same $\infty$-jet at a given point.
}

\vspace{3mm}
Jets have been introduced by mathematicians in order to make sense of locality (in the sense of classical field theory) without necessarily going into the details of functional space analysis. Indeed, the $k$-jet spaces of finite order $k$ are finite-dimensional. Therefore, jet spaces allow to study (some aspects of) functions on a manifold
via the basic tools of abstract algebra. Another motivation for introducing jets is that they allow to provide a coordinate-independent definition of higher-order partial differential equations.

\vspace{3mm}
\noindent{\small\textbf{Remark:}  Note that the $k$-th derivative alone 
is \textit{not} a coordinate-free object, only the collection of all its derivatives till order $k$ is a geometric object
(since the coefficients $\partial_{\mu_1}\ldots\partial_{\mu_r} f(x)$ do \textit{not} transform as symmetric covariant tensor fields of rank $r$ but are mixed under general coordinate transformations).
However, as can be expected from the coordinate expression and as can be checked from the transformation law \eqref{coordtransfojet} under coordinate changes,
the space of symmetric covariant tensors of rank $k$ is isomorphic to the quotient of the contact ideal of order $k-1$ by the contact ideal of order $k$, \textit{cf.} \eqref{cotangentcontactideals}. 
In other words, the space of symmetric covariant tensors is the cokernel (\textit{i.e.} the quotient of the codomain by the image) of the canonical inclusions \eqref{icontactid}.
}

\subsection{Higher-order cojets}

As will become clear, the algebraic definition of differential operators via commutators is equivalent to another possible definition of differential operators insisting on locality: differential operators are those operators on the vector space $ C^\infty(M)$ that are \textit{local} in the sense that their action at a point only depends on the smooth structure (\textit{i.e.} the jet) of the given function at that point (in other words, the contact ideal is annihilated). To be more precise, the restriction $\hat{X}|_m$ of a differential operator of order $k$ at a single point $m\in M$ is a linear form on the $k$-jet space $J^k_mM$. This motivates the introduction of some specific terminology for the values of differential operators at a given point: cojets.

\subsubsection{Higher-order cojets as generalised functions}

The dual of the $k$th jet space, \textit{i.e.} the space 
\be
D^{k}_mM\,:=\,(J^k_mM)^*\,,\qquad (k\in\mathbb{N})
\ee
of linear forms on the $k$-jet space $J^k_mM$, will be called the space of $k$-\textbf{cojets} at $m$. 
Equivalently, a $k$-cojet at $m$ can be thought as the \textbf{value at the point} $m$ \textbf{of a differential operator} $\hat{X}\in\mathcal{D}^k(M)$ of order $k$, defined as the linear form 
\be
\hat{X}|_m:=\delta_m\circ\hat{X}
\ee
on $C^\infty(M)$, where $\delta_m:C^\infty(M)\to\mathbb R$ is the evaluation functional at the point $m$.
Since the contact ideal $\mathcal{I}^k(m)$ of order $k$ at the point $m$ belongs to the kernel of any such linear form $\hat{X}|_m:C^\infty(M)\to\mathbb R$, one may indeed consider that the latter effectively defines a linear form on the $k$-jet space at $m$, \textit{c.f.} \eqref{JkM}.
Furthermore, a $k$-cojet at $m$ can be also be defined as an equivalence classes of differential operators of order $k$, where two differential operators $\hat{X}$ and $\hat{Y}$ are equivalent if they produce the same result, at the point $m$, on any given function. This last definition agrees with the previous one, due to the relation \eqref{equivreltgtvect}.

\vspace{5mm}\begin{figure}[h!]
\begin{framed}
\begin{center}
\textbf{Equivalent formulations of finite-order cojets}
\end{center}

\noindent
On a smooth manifold, the following notions are equivalent for any integer $k\in\mathbb{N}$:
\begin{enumerate}
 \item a $k$-cojet at a point $m$,
 \item a linear form on the $k$-jet space $J^k_mM$,
 \item the value $\hat{X}|_m$ at $m$ of a differential operator $\hat{X}\in\mathcal{D}^k(M)$ of order $k$.
\end{enumerate}
\vspace{3mm}
\end{framed}
\end{figure}

For $k=\infty$, the above definitions require more care. The topological dual of the $\infty$-jet space, \textit{i.e.} the space 
\be
D_mM\,:=\,(J^\infty_mM)^\prime\,.
\ee
of continuous linear forms on the $\infty$-jet space $J^\infty_mM$, will be called the space of $\infty$-\textbf{cojets} at $m$.
In the language of distribution theory, a smooth function with compact support is called a \textbf{test function}.
The subspace of test functions is sometimes denoted $C_c^\infty(M)\subset C^\infty(M)$.
 One may consider its topological dual $C_c^\infty(M)^\prime$ spanned by the continuous linear 
functionals on $C_c^\infty(M)$. These functionals are called \textbf{generalised functions} \cite{GelfandShilov} (or ``distributions'' \cite{Schwartz}) on the manifold $M$. 
The space of $\infty$-cojets at $m$ can be defined as the space of generalised functions on $M$ whose support is the point $m$.
A theorem of Schwartz states that a generalised function whose support is a single point decomposes a finite linear combination of derivatives of the Dirac distribution at this point \cite[Th.35,Chap.3]{Schwartz}.
For instance, remember that the space $D^0_mM$ of $0$-cojets at $m$, \textit{i.e.} the vector space dual to $J^0_mM= C^\infty(M)/{\mathcal I}^0(m)\cong \mathbb R$, can be thought of as the one-dimensional space spanned by the evaluation functional $\delta_m$. More generally, the space $D^{k}_mM$ of $k$-cojets at $m$ is spanned by all  derivatives of the Dirac distribution at $m$ till order $k$.
Accordingly, in a local coordinate system a possible representation for a $k$-cojet $X$ at a point $m\in M$ of coordinates $y^\mu$ is as a generalised function: 
\be
X(x) =\sum\limits_{r=0}^k \frac1{r!}\,X^{\mu_1\ldots \mu_r}(y)\,\frac{\partial}{\partial x^{\mu_1}}\ldots \frac{\partial}{\partial x^{\mu_r}}\delta^n(y-x)\,,
\ee
Indeed, via this realisation, the $k$-cojet acts on a test function $\phi$ as follows:
\be
\langle X, \phi\rangle =\int dx\, X(x)\,\phi(x) = \sum\limits_{r=0}^k \frac1{r!}\,X^{\mu_1\ldots \mu_r}(y)\,\partial_{\mu_1}\cdots\partial_{\mu_r}\phi(y)\,.
\ee

In a more algebraic language, a compact expression for a $k$-cojet $X$ is as a polynomial of degree $k$
in the auxiliary variable $\partial_\varepsilon$: 
\be
X(\partial_\varepsilon) =\sum\limits_{r=0}^k \frac1{r!}\,X^{\mu_1\ldots \mu_r}\,\frac{\partial}{\partial \varepsilon^{\mu_1}}\ldots \frac{\partial}{\partial \varepsilon^{\mu_r}}\,,
\ee
where the location of the cojet was not specified. 
Indeed, via this choice of notation, the $k$-cojet $ X(\partial_\varepsilon)$ above acts on $k$-jets $\phi(\varepsilon)$, realised as truncated polynomials, like a differential operator 
of order $k$ (for the auxilliary coordinate $\varepsilon$) followed by an evaluation at $\varepsilon=0$:
\be
\langle X, \phi\rangle =\left. X(\partial_\varepsilon)\phi(\varepsilon)\,\right|_{\varepsilon=0} = \sum\limits_{r=0}^k \frac1{r!}\,X^{\mu_1\ldots \mu_r}\,\phi_{\mu_1\ldots \mu_r}\,.
\ee

\vspace{5mm}\begin{figure}[h!]
\begin{framed}
\begin{center}
\textbf{Equivalent formulations of cojets}
\end{center}

\noindent
On a smooth manifold, the following notions are equivalent:
\begin{enumerate}
 \item a cojet at a point $m$,
 \item a continuous linear form on the $\infty$-jet space $J^\infty_mM$,
 \item the value $\hat{X}|_m$ at $m$ of a differential operator $\hat{X}\in\mathcal{D}(M)$,
 \item a generalised function with support at $m$.
\end{enumerate}
\vspace{3mm}
\end{framed}
\end{figure}

\subsubsection{Higher-order cojet bundles}

The $k$-\textbf{cojet bundle} $D^kM=\bigcup_m D^k_mM$ has local coordinates 
$$(x^\nu,X,X^\mu,X^{\mu_1\mu_2},\ldots,X^{\mu_1 \cdots\,\mu_k}).$$
The transformation law of a $k$-cojet under the coordinate transformation $x^\mu\mapsto x^{\prime\mu}(x)$ is quite complicated for high  $k$, so we only give the transformation law of
2-cojets:
\be
X^\prime=X\,,\quad X^{\prime\mu}=\frac{\partial x^{\prime\mu}}{\partial x^{\nu}}\,X^\nu
+\frac{\partial^2 x^{\prime\mu}}{\partial x^{\nu_1}\partial x^{\nu_2}}\,X^{\nu_1\nu_2}\,,\quad 
X^{\prime\mu_1\mu_2}=\frac{\partial x^{\prime\mu_1}}{\partial x^{\nu_1}}\,\frac{\partial x^{\prime\mu_2}}{\partial x^{\nu_2}}
\,X^{\nu_1\nu_2}\,.
\label{coordtransfocojet}
\ee
More generally, all (but only) the higher rank components contribute in the transformation law of a given component of a $k$-cojet, thus its lower components $X^{\mu_1 \cdots\,\mu_r}$ of rank $0<r<k$ do \textit{not} transform as rank-$r$ symmetric contravariant tensors.
In other words, only the collection of all its non-trivial components $X^\mu$, $X^{\mu_1\mu_2}$, ..., $X^{\mu_1 \cdots\,\mu_k}$ is
a single coordinate-free object. More accurately, only the rank-zero and rank-$k$ components of a $k$-cojet can be separated in a coordinate-invariant way.
An important motivation behind jet theory is to allow a geometric definition of differential operators.
Contrarily to tangent and cotangent vectors, their higher-order generalisation (cojets and jets, respectively) do not have nice transformation properties under coordinate transformations, so their transformation laws have not been written down here in details although a tensor-like calculus can be designed to handle them \cite{Foster}. 
If the components of metric-like fields are, say cojets (as suggested by the higher-spin Lie derivative), then they do \textit{not} transform as symmetric tensor fields under diffeomorphisms. Such property might be at the root of the well-known obstruction in higher-spin gravity to ``naive'' minimal coupling of higher-spin particles to gravitons (see \textit{e.g.} \cite{Introductions} and refs therein).

The sections of the $k$-cojet bundle  $D^kM$ will be called
$k$-\textbf{cojet fields}.
Locally, the latter are compactly expressed as generating functions
$X(x;\partial_\varepsilon)$ which are smooth in the coordinates $x^\mu$ and polynomial in the auxiliary variable $\partial/\partial\varepsilon^\nu$.
There is an infinite sequence of embeddings (actually, an $\mathbb N$-filtration) of vector bundles on $M$:
\be
M\times {\mathbb R}^*\cong D^0M\hookrightarrow D^1M \hookrightarrow D^2M \hookrightarrow \ldots \hookrightarrow D^kM \hookrightarrow D^{k+1}M \hookrightarrow \ldots
\ee
the direct limit of which is the infinite-order cojet bundle $DM$ that will be called the \textbf{enveloping bundle} of $M$.\footnote{This term is motivated by the importance of universal enveloping algebras of finite-dimensional Lie algebras in higher-spin gravity. Moreover, this terminology is in line with the fact that the almost-commutative algebra of differential operators ${\mathcal D}(M)$ is the universal enveloping algebra of the Lie-Rinehart algebra of vector fields ${\mathfrak{X}}(M)$.}
Accordingly, the infinite-order cojet space $D_mM$ will be called  the \textbf{enveloping space} at $m\in M$.

\vspace{3mm}
\noindent{\small\textbf{Example (jet vs cojet space)\,:} The $\infty$-cojet space $D_0V\cong \odot(V)$ at the origin of the vector space $V$ is the topological dual of the $\infty$-jet space $J^\infty_0 V\cong \overline{\odot}(V^*)$. Conversely, the $\infty$-jet space $J^\infty_0 V$ is the algebraic dual of the $\infty$-cojet space $D_0V$.}
\vspace{3mm}

The cokernel of the $k$th embedding $D^{k-1}M\hookrightarrow D^{k}M$ is the bundle of symmetric contravariant tensors of rank $k$, i.e. $\odot^kTM$. In other words, the quotient of $k$-cojet space by the ($k-1$)-cojet space is isomorphic to the space of symmetric contravariant tangent tensors of rank $k$: 
\be
\odot^kTM\,\cong\, D^{k}M\,/\,D^{k-1}M\,.
\ee
A representative of the equivalence class of a $k$-cojet is its principal symbol. 

\subsubsection{Differential operators as cojet fields}

Differential operators of order $k$ can be defined as linear operators $\hat{X}: C^\infty(M)\to  C^\infty(M)$ 
which factor through the $k$-jet bundle 
$J^kM$ in the sense that $\hat{X}={X}\circ j^k$ where ${X}\in \Gamma(D^{k}M)$ is a $k$-cojet field, \textit{i.e.} a section of the $k$-cojet bundle $D^{k}M$, and $j^k: C^\infty(M)\to\mathcal{J}^k(M)$ is the order-$k$ prolongation map.
In local coordinates, the correspondence goes as follows:
\begin{eqnarray}
(\hat{X}f)(x)&=& \left.\left[\, X(x;\partial_\varepsilon)\,(j^kf)(x;\varepsilon)\,\right]\right|_{\varepsilon=0}\nonumber\\
&=&\Big( \,X\big(j^kf\big)\,\Big)(x)\,, 
\end{eqnarray}
which can be checked by making use of the previous explicit local expressions for the differential operator $\hat{X}: C^\infty(M)\to  C^\infty(M)$ of order $k$, 
its $k$-cojet field $X:\mathcal{J}^k(M)\to  C^\infty(M)$ and the prolongation map $j^k: C^\infty(M)\to\mathcal{J}^k(M)$.
Due to this equivalent definition of differential operators, we will 
often identify differential operators with cojet fields. 
Actually, the identification corresponds to the formal replacement 
$\partial_\varepsilon$ with $\partial_x$ together with the normal ordering prescription (related to the previous subtlety in the isomorphism).

\vspace{5mm}\begin{figure}[h!]
\begin{framed}
\begin{center}
\textbf{The many faces of differential operators}
\end{center}

\noindent
Given a smooth manifold $M$, the following notions are equivalent:
\begin{enumerate}
 \item a linear scalar-valued differential operator $\hat{X}\in{\mathcal D}^k(M)$ of order $k$ on $M$,
 \item an endomorphism $\hat{X}\in\text{End}\big(C^\infty(M)\big)$ of the vector space $C^\infty(M)$ such that $[\,[\,\ldots\,[\hat{X}\,,\,\hat{f}_1]\,,\,\hat{f}_2]\ldots \,,\,\hat{f}_k]\in{\mathcal D}^0(M)$ for any functions $f_1,f_2,\cdots,f_k\in C^\infty(M)$,
 \item an endomorphism $\hat{X}\in\text{End}\big(C^\infty(M)\big)$ of the vector space $C^\infty(M)$ which is almost $C^\infty(M)$-linear in the sense that $[\hat{X},\hat{f}]\in {\mathcal D}^{k-1}(M)$ for any function $f\in C^\infty(M)$,
 \item a $k$-cojet field $X\in\Gamma(D^kM)$ on $M$, \textit{i.e.} a section of the $k$-cojet bundle $D^{k}M$.
\end{enumerate}
\vspace{3mm}
\end{framed}
\end{figure}

\subsection{Pushforward of differential operators and cojets}

Let $F:M\to N$ be a map from the manifold $M$ (source) to the manifold $N$ (target). The following definitions are the direct higher-order generalisations of their standard versions for vector fields.

Two differential operators $\hat{X}\in {\mathcal D}(M)$ and $\hat{Y}\in {\mathcal D}(N)$ are said to be \textbf{related by} $F$ if 
\be
\hat{X}\circ F^*=F^*\circ \hat{Y}\label{related}
\ee 
where $F^*\,:\, C^\infty(N)\to  C^\infty(M)$ is the pullback of $F:M\to N$. The equality \eqref{related} makes sense since the differential operators $\hat{X}$ and $\hat{Y}$ are seen as endomorphisms of $ C^\infty(M)$ and $ C^\infty(N)$ respectively while $F^*$ is a homomorphism from $ C^\infty(N)$ to $ C^\infty(M)$.
More explicitly, the condition \eqref{related} reads as follows:
\be\label{related2}
\hat{X}\big[F^*(g)\big]=F^*\big(\hat{Y}[g]\big)\,,\qquad\forall g\in C^\infty(N)\,.
\ee
This condition is sufficiently subtle and important to deserve a third equivalent writing: 
\be
\hat{X}[\,g\circ F]=\hat{Y}[g]\circ F\,, \qquad\forall g\in C^\infty(N)\,.
\ee

If $F$ is bijective then any differential operator $\hat{X}$ on the source $M$ is related by $F$ to a unique differential operator $\hat{Y}$ on the target $N$, which is called the \textbf{pushforward of} $\hat{X}$ by $F$ and the corresponding $\hat{Y}$ in \eqref{related} is denoted $F_*\hat{X}$.
Therefore, the 
\textbf{pushforward by a bijective} $F$ is defined as the algebra isomorphism
\be
F_*\,:\,{\mathcal D}(M)\stackrel{\sim}{\to} {\mathcal D}(N)\,:\,\hat{X}\mapsto (F^{-1})^*\circ\hat{X}\circ F^*\,.
\ee
The pushforward preserves the filtration by the order of the differential operators.
If $F$ is injective then its restriction is invertible when the codomain is restricted to $F(M)\subseteq N$. Therefore, the 
\textbf{pushforward by an injective} $F:M\hookrightarrow N$ is a well defined map $F_*\,:\,{\mathcal D}\big(M\big)\hookrightarrow {\mathcal D}\big(F(M)\big)$.
However, if $F$ is surjective, then the pushforward is not always well defined, which motivates the following definition:
when a differential operator $\hat{X}$ on $M$ is related by a surjective $F$ to a (well-defined) differential operator $\hat{Y}$ on $N$,
then the former $\hat{X}$ is called \textbf{projectable on} $N$ (by the pushforward $F_*$) while the latter $\hat{Y}$ is called the pushforward of $\hat{X}$ by $F$ and is denoted $F_*\hat{X}$($=\hat{Y}$).
The space of differential operators on $M$ projectable on $N$ by $F$ will be denoted as the preimage $(F_*)^{-1}{\mathcal D}(N)$. In this way, it becomes tautological to say that the \textbf{pushforward by a surjective} $F:M\twoheadrightarrow N$ is a well defined map 
$F_*\,:\,(F_*)^{-1}{\mathcal D}(N)\twoheadrightarrow {\mathcal D}(N)$.

All these definitions of pushforward by $F$ are morphisms of algebras, \textit{i.e.} $F_*(\hat{X}_1\circ \hat{X}_2)=
(F_*\hat{X}_1)\circ(F_*\hat{X}_2)$ for any differential operators $\hat{X}_1$ and $\hat{X}_2$ on the source $M$.
Therefore the kernel, $\text{Ker}\,F_*$, of the pushforward by a surjective map $F$ is an associative ideal of the algebra of projectable differential operators. The corresponding quotient is isomorphic to the algebra of differential operators on the target manifold: 
${\mathcal D}(N)\cong (F_*)^{-1}{\mathcal D}(N)\,/\,\text{Ker}\,F_*$.
 
The pushforward map itself $F\mapsto F_*$ is a homomorphism from the associative algebra of smooth maps between manifolds to the associative algebra of morphisms between associative algebras of differential operators, \textit{i.e.} it preserves the order of multiplication: $(F\circ\, G)_*=F_*\circ\, G_*$. 

In another generalisation of Milnor exercise by Grabowski and Poncin, the commutative algebra of functions is replaced with the associative\footnote{Actually, the theorem holds for the weaker structure of \textit{Lie} (rather than \textit{associative}) algebra of differential operators \cite{Grabowski:2002}.} algebra of differential operators \cite{Grabowski:2002}. 

\vspace{3mm}
\noindent{\textbf{Theorem (Grabowski \& Poncin)\,:} \textit{A map $\Phi:{\mathcal D}(M)\stackrel{\sim}{\to}{\mathcal D}(N)$ between the algebras of differential operators on two smooth manifolds is an isomorphism of associative algebras iff it is the pushforward of a diffeomorphism $F:M\stackrel{\sim}{\to}N$ between these two manifolds, \textit{i.e.} $\Phi=F_*$.}}
\vspace{3mm}

The essence of the proof relies on showing that any such isomorphism of associative algebras is also an isomorphism of filtered algebra (in which case, one can reduce the problem to the genuine Milnor exercise).
The various Milnor exercises that were reviewed till now gives a mathematically precise meaning to the equivalences \eqref{equivalence} pictured in Section \ref{motivation}:

\vspace{2mm}
Consider a differential operator $\hat{X}$ on $M$ related by $F:M\to N$ to the differential operator $F_*\hat{X}$ on $N$, \textit{cf.} \eqref{related2}.
For any function $g$ on $N$, the functions $\hat{X}[F^*g]$ and $F_*\hat{X}[g]$ are functions on $M$ and $N$ respectively.
Moreover, the evaluation at the point $m\in M$ of the function $\hat{X}[F^*g]$
is equal to the evaluation at the point $F(m)\in N$ of the function $F_*\hat{X}[g]$.
In other words, we have the familiar equality
\be
\hat{X}[F^*g]|_m\,=\,F_*\hat{X}[g]|_{F(m)}.
\label{evaluationequal}
\ee
A $k$-cojet $X_m$ at a point $m\in M$
can be identified with the equivalence class of differential operators $\hat{X}$ on $M$ of order $k$ with the same value $\hat{X}|_m=\delta_m\circ \hat{X}$ at $m$.
The remarkable fact about the equality \eqref{evaluationequal} is that it only involves values at a single point of the relevant objects (the $k$-cojet $X_m$ and the $k$-jet $j^k_mf$).
Therefore all the previous problems for pushforwards of cojet fields do not arise for individual cojets, because the latter are only defined at a single point. In particular,
the pushforward of a cojet by a map $F:M\to N$
is well-defined independently of the injectivity/surjectivity properties of the map $F$. In fact, the following definition of
the pushforward 
\be
F_*\hat{X}|_m\,:=\,\hat{X}|_m\circ F^*
\ee
of a $k$-cojet $\hat{X}|_m$ at a point $m$ agrees with \eqref{evaluationequal}.

The pushforward by $F:M\to N$ is the vector bundle morphism $F_*\,:\,D^kM\to D^kN$ corresponding in the fibre to the linear maps
\be
F_{*m}\,:\,D^k_mM\to D^k_{F(m)}N\,:\,X_m \mapsto (F_*X)_{F(m)}\,,
\ee
at each point.
Actually this notion of pushforward ensures that the map $D:M\mapsto DM$ sending a manifold to its enveloping bundle defines a covariant functor from the category of smooth manifolds to the category of vector bundles, where any smooth map $F:M\to N$ between two manifolds is sent to the morphism $F_*\,:\,DM\to DN$ of vector bundles. In the language of category theory, the pushforward $F_*\,:\,DM\to DN$ would be denoted as $DF$. The map $D:M\mapsto DM$ sending a manifold to its enveloping bundle also defines a covariant functor from the category of smooth manifolds to the category of vector bundles.   
In the first-order case $k=1$, one may restrict to the tangent bundle, in which case the pushforward 
by $F:M\to N$ is the vector bundle morphism $TF\,:\,TM\to TN$ also known as the \textbf{differential of} $F$.

\pagebreak

\section*{Acknowledgments}

I am grateful to Thomas Basile for his patient reading and helpful comments on some early version of these notes.



\begin{thebibliography}{99}

\bibitem{Reviews}
	M.~A.~Vasiliev,
	``Higher spin gauge theories in four-dimensions, three-dimensions, and two-dimensions,''
  Int.\ J.\ Mod.\ Phys.\ D {\bf 5} (1996) 763
  [hep-th/9611024];
  ``Higher spin gauge theories in various dimensions,''
  Fortsch.\ Phys.\  {\bf 52} (2004) 702
  [hep-th/0401177];
	``Higher spin gauge theories in any dimension,''
  Comptes Rendus Physique {\bf 5} (2004) 1101
  [hep-th/0409260];\\
	X.~Bekaert, S.~Cnockaert, C.~Iazeolla and M.~A.~Vasiliev,
  ``Nonlinear higher spin theories in various dimensions,''
  hep-th/0503128;\\
	V.~E.~Didenko and E.~D.~Skvortsov,
  ``Elements of Vasiliev theory,'' arXiv:1401.2975 [hep-th];\\
	M.~A.~Vasiliev,
	``Higher-spin theory and space-time metamorphoses,''
  Lect.\ Notes Phys.\ {\bf 892} (2015) 227
  [arXiv:1404.1948 [hep-th]];\\
	D.~Ponomarev,
	``Basic introduction to higher-spin theories,''
	arXiv:2206.15385 [hep-th].

\bibitem{Introductions}
	D.~Sorokin,
  ``Introduction to the classical theory of higher spins,''
  AIP Conf.\ Proc.\  {\bf 767} (2005) 172
  [hep-th/0405069];\\
  X.~Bekaert, N.~Boulanger and P.~Sundell,
  ``How higher-spin gravity surpasses the spin two barrier: no-go theorems versus yes-go examples,''
  Rev.\ Mod.\ Phys.\  {\bf 84} (2012) 987
  [arXiv:1007.0435 [hep-th]];\\
	R.~Rahman,
	``Higher Spin Theory - Part I,''
	PoS \textbf{ModaveVIII} (2012) 004
	[arXiv:1307.3199 [hep-th]];\\
	R.~Rahman and M.~Taronna,
  ``From Higher Spins to Strings: A Primer,''
  arXiv:1512.07932 [hep-th];\\
	P.~Kessel,
  ``The Very Basics of Higher-Spin Theory,''
  PoS \textbf{Modave2016} (2017) 001
  [arXiv:1702.03694 [hep-th]];\\
	A.~Bengtsson, \textit{Higher Spin Field Theory (Concepts, Methods and History) Volume 1: Free Theory} (De Gruyter, 2020);\\
	X.~Bekaert, N.~Boulanger, A.~Campoleoni, M.~Chiodaroli, D.~Francia, M.~Grigoriev, E.~Sezgin and E.~Skvortsov,
	``Snowmass White Paper: Higher Spin Gravity and Higher Spin Symmetry,''
	arXiv:2205.01567 [hep-th].
	
\bibitem{Proceedings}
	R.~Argurio, G.~Barnich, G.~Bonelli and M.~Grigoriev (eds), \textit{Higher Spin Gauge Theories} (International Solvay Institutes, 2004);\\
	L.~Brink, M.~Henneaux and M.~A.~Vasiliev (eds), \textit{Higher Spin Gauge Theories} (World Scientific, 2017).
	
\bibitem{Saunders}
D.~J.~Saunders, \textit{The geometry of jet bundles}  (Cambridge University Press, 1989).

\bibitem{Olver}
P.~J.~Olver, \textit{Equivalence, Invariants and Symmetry} (Cambridge University Press, 1995)

\bibitem{Krasilshchik}
I.~S.~Krasilshchik and A.~M.~Vinogradov (Eds),	\textit{Symmetries and Conservation Laws for Differential Equations of Mathematical Physics} (American Mathematical Society, 1999).

\bibitem{Sardanashvily1}
G.~Sardanashvily,	\textit{Advanced Differential Geometry for Theoreticians: Fiber Bundles, Jet Manifolds and Lagrangian Theory} (Lambert Academic Publishing, 2013) [arXiv:0908.1886 [math-ph]].

\bibitem{Vinogradov}
A.~M.~Vinogradov, ``An informal introduction to the geometry of jet spaces,'' Rendiconti Seminari Facolt\`a Scienze Universit\`a Cagliari (supplemento al volume) {\bf 58} (1988); ``Introduction to Secondary Calculus,'' Diffiety Institute Preprint Series 5/98;\\
G.~Sardanashvily,	``Five lectures on jet manifold methods in field theory,'' arXiv:hep-th/9411089; ``Ten lectures on jet manifolds in classical and quantum field theory,'' arXiv:math-ph/0203040.

\bibitem{Nestruev}
J.~Nestruev, \textit{Smooth Manifolds and Observables}  (Springer, 2003).

\bibitem{Sardanashvily2}
G.~Sardanashvily, \textit{Lectures on differential geometry of modules and rings: Application to Quantum Theory} (Lambert Academic Publishing, 2012) [arXiv:0910.1515 [math-ph]].

\bibitem{Atiyah}
M.~Atiyah and I.~G.~MacDonald, \textit{Introduction To Commutative Algebra} (Westview Press, 1994).

\bibitem{Milnor}
J.~Milnor and J.~D.~Stasheff,	\textit{Characteristic Classes}, Annals of Mathematics Studies {\bf 76} (Princeton University Press, 1974).

\bibitem{Kolar}
I.~Kolar, J.~Slovak and P.~W.~Michor, \textit{Natural operations in differential geometry} (Springer-Verlag, 1993).

\bibitem{Grabowski:2003a}
J.~Grabowski, ``Isomorphisms of algebras of smooth functions revisited,'' J.\ Arch.\ Math.\ {\bf 85}  (2005) 190 [arXiv:math/0310295 [math.DG]]. 

\bibitem{Lee}
J.~M.~Lee, \textit{Introduction to Smooth Manifolds}, Graduate Texts in Mathematics {\bf 218} (1st edition: Springer, 2003) Chapter 17; (2nd edition: Springer, 2012) Chapter 9.

\bibitem{Takens}
F.~Takens, ``Derivations of vector fields,'' Compos. Math. {\bf 26} (1973) 151.

\bibitem{Pursell}
M.~E.~Shanks and L.~E.~Pursell, ``The Lie Algebra of a Smooth Manifold,'' Proc. Am. Math. Soc. {\bf 5} (1954) 468.

\bibitem{DuboisV}
M.~Dubois-Violette and P.~W.~Michor, ``A common generalization of the Frohlicher-Nijenhuis bracket 
and the Schouten bracket for symmetric multivector fields'', Indag.\ Mathem.\ N.S.\ {\bf 6}  (1995) 51.

\bibitem{Grabowski:2003b}
J.~Grabowski and N.~Poncin, ``Derivations of the Lie algebras of differential operators'', Indag.\ Mathem.\ N.S.\ {\bf 16}  (2005) 181 [ 	arXiv:math/0312162 [math.DG]].

\bibitem{Grabowski:2002}
J.~Grabowski and N.~Poncin, ``Automorphisms of quantum and classical Poisson algebras'', Compositio Math.\ {\bf 140}  (2004) 511 [arXiv:math/0211175 [math.RA]].

\bibitem{Grothendieck:1967}
A.~Grothendieck, ``\'El\'ements de g\'eom\'etrie alg\'ebrique : IV. \'Etude locale des sch\'emas et des morphismes de sch\'emas, Quatri\`eme partie,''  Publ.\ Math.\ IH\'ES {\bf 32}  (1967) 5.

\bibitem{Bekaert:2021sfc}
X.~Bekaert, ``Notes on Higher-Spin Diffeomorphisms,'' Universe \textbf{7} (2021) 508 [arXiv:2108.09263 [hep-th]].

\bibitem{KanelBelov:2005}
A.~Kanel-Belov and M.~Kontsevich, ``Automorphisms of Weyl algebras,''  Lett.\ Math.\ Phys.\ {\bf 74}  (2005) 181 [arXiv:math/0512169 [math.RA]].

\bibitem{KanelBelov:2018}
A.~Kanel-Belov, A.~Elishev and J.~T.~Yu, ``Automorphisms of Weyl Algebra and a Conjecture of Kontsevich,'' arXiv:1802.01225
[math.RA].

\bibitem{Dixmier:1968}
J.~Dixmier,``Sur les alg\`ebres de Weyl,'' Bull.\ Soc.\ Math.\ France {\bf 96}  (1968) 209. 

\bibitem{Adjamagbo}
K.~Adjamagbo and A.~van den Essen, ``On the equivalence of the Jacobian, Dixmier and Poisson Conjectures in any characteristic,'' arXiv:math/0608009.

\bibitem{Borel}
\'E.~Borel, ``Sur quelques points de la th\'eorie des fonctions,'' Ann. Sci. de l’\'Ecole Normale Sup\'erieure (S\'erie III) {\bf 12} (1895) 9.

\bibitem{Weil}
A.~Weil, ``Th\'eorie des points proches sur les vari\'et\'es diff\'erentielles,'' Colloq.\ Internat.\ Centre Nat.\ Rech.\ Sci.\ {\bf 52}  (1953) 111.

\bibitem{GelfandShilov}
I.~M.~Gelfand and G.~E.~Shilov, \textit{Generalized functions}, volumes 1–5 (Academic Press, 1966–1968).

\bibitem{Schwartz}
L.~Schwartz, \textit{Th\'eorie des distributions}, tomes I-II (Hermann, 1950-1951).

\bibitem{Foster}
B.~L.~Foster, ``Differentiation on manifolds without a connection,'' Michigan Math.\ J.\ {\bf 5}  (1958) 183;
``Would Leibniz lie to you?'', Math.\ Intelligencer {\bf 8}  (1986) 34; see also the list of references in:\\
O.~E.~Barndorff-Nielsen and P.~Blaesild, ``Coordinate-free definition of structurally symmetric derivative string'', Adv.\ App.\ Math.\ {\bf 9}  (1988) 1.\footnote{The translation rule from the terminology of the latter two authors to ours is as follows: `derivative string' $\mapsto$ jet field, `differentiation string' $\mapsto$ cojet field.}

\end{thebibliography}
\end{document}